\documentclass{article}
\usepackage{fullpage}
\usepackage{graphicx}
\usepackage{amsmath}
\usepackage{amssymb}
\title{A Variable Flavour Number Scheme for Heavy Quark Production at small $x$}
\date{}
\renewcommand{\vec}[1]{\mbox{\boldmath$ #1 $}}
\newcommand{\mat}[1]{\ensuremath{\bf{\textsf{#1}}}}
\begin{document}
\bibliographystyle{utphys}
\newcommand{\msbar}{\ensuremath{\overline{\text{MS}}}}
\newcommand{\DIS}{\ensuremath{\text{DIS}}}
\newcommand{\abar}{\ensuremath{\bar{\alpha}_S}}
\newcommand{\bb}{\ensuremath{\bar{\beta}_0}}
\setlength{\parindent}{0pt}

\titlepage
\begin{flushright}
Cavendish-HEP-2006/08 \\
\end{flushright}

\vspace*{0.5cm}

\begin{center}
{\Large \bf A Variable Flavour Number Scheme for Heavy Quark Production at Small $x$ }

\vspace*{1cm}
\textsc{C.D. White$^{a,}$\footnote{cdw24@hep.phy.cam.ac.uk} and R.S. Thorne$^{a,}$\footnote{thorne@hep.phy.cam.ac.uk}} \\

\vspace*{0.5cm} $^a$ Cavendish Laboratory, University of Cambridge, \\ J J Thomson Avenue,
Cambridge, CB3 0HE, UK
\end{center}

\vspace*{0.5cm}

\begin{abstract}
We define a new variable flavour number scheme for use in deep inelastic scattering, motivated by the need to consistently implement high energy resummations alongside a fixed order QCD expansion. We define the ${\DIS}(\chi)$ scheme at fixed order, and show how to obtain the small $x$ coefficient functions and heavy flavour matrix elements to leading order in the high energy resummation. We then implement these results in a global fit at LO which includes leading resummations with running coupling corrections. Finally, we address the impact of the resummed results on predictions for the longitudinal structure function. We find that they stabilise the behaviour of $F_L$ at small $x$. Overall, we find that resummations significantly improve the fit to scattering data in the low $x$ regime, although higher orders in the fixed order expansion are needed to describe current structure function and related data over the complete $x$ range.
\end{abstract}

\vspace*{0.5cm}

\section{Introduction}
The high partonic centre of mass energies available at current and forthcoming particle collider experiments impart a greater phenomenological significance to the production of heavy quark flavours. In deep inelastic scattering (DIS), for example, heavy flavours can be produced by boson gluon fusion \cite{Witten} once the partonic centre of mass energy $\hat{W}^2$ reaches $4M^2$, where $M$ is the mass of the heavy quark. For momentum transfers $Q$ satisfying $Q^2\gg M^2$, one may take the heavy flavours into account by defining heavy parton distributions obeying the massless evolution equations \cite{ZMF}. When $Q^2$ is comparable with $M^2$, however, a consistent treatment of threshold effects must be implemented. Such descriptions have been around for a long time as so called general mass variable flavour number schemes \cite{ACOT1,ACOT,Buza2,RT,RT2,Chuvakin}, and have been refined alongside the onset of data on the proton structure functions at higher $Q^2$ \cite{F2c1,ZEUSb,F2c2,H1c,H1b,H1a,ZEUSa}.\\

Another problem encountered at high energy is that the coefficient functions relating the proton structure functions to the parton distribution functions depend on the logarithm of the Bjorken $x$ variable. So also do the splitting functions which govern the evolution of the partons. At sufficiently low $x$ it may be necessary to resum $\log{1/x}$ terms, thus supplementing the traditional perturbation series ordered in fixed powers of $\alpha_S$. There is already evidence that such a resummation is necessary in order to improve QCD fits to scattering data at fixed order. For example, NNLO fits seem to be improved by the addition of higher order terms involving phenomenological small $x$ logarithms \cite{MRST_errors}, whose coefficients are determined by the data. The longitudinal structure function obtained from the reduced cross-section measured at HERA appears to be inconsistent with the theoretical prediction using NLO QCD at small $x$ \cite{ThorneFL}, indicating the importance of higher order contributions. Results for the $\msbar$ scheme three-loop coefficient functions for $F_2$ \cite{Vogtcoeffs2} and $F_L$ \cite{VogtcoeffsL} demonstrate an apparent perturbative instability in both quantities at low $x$. In order to resum the high energy contributions one uses the BFKL equation \cite{BFKL}, an integral equation for the unintegrated gluon 4-point function whose kernel is presently known to NLL order \cite{Fadin,CamiciNLL}.\\

The aim of this paper is to present a scheme for the description of heavy quark flavours in DIS, applicable to both the small $x$ and fixed order expansions. We implement our scheme in a global fit to scattering data at LO which, in the description of $F_2$, incorporates small $x$ resummations at leading logarithmic (LL) order with running coupling corrections. We also present resummed predictions for the longitudinal structure function $F_L$, which at low $x$ is a sensitive probe of the gluon distribution. At high inelasticity $y$, one cannot measure $F_2$ directly but instead obtains a combination of $F_2$ and $F_L$ (the reduced cross-section). It is therefore necessary to have good theoretical predictions for the longitudinal structure function. Furthermore, it is widely hoped that measurements of $F_L$ will discriminate between different low $x$ models \cite{Partons, ThorneFL}.\\

The structure of the paper is as follows. In section 2 we briefly summarise variable flavour number schemes to introduce our notation, before outlining our proposed scheme (the $\DIS(\chi)$ scheme) in the fixed order expansion and at small $x$ to LL order. In section 3 we present our implementation of the small $x$ resummation adopting the method of \cite{Thorne01}, which includes running coupling corrections. In section 4 we present the results of a global fit at LO which includes the resummed heavy flavour contributions. We discuss how this differs from a NLO QCD fit. Finally, in section 5, we discuss our resummed predictions for the longitudinal structure function. 

\section{The $\DIS(\chi)$ Scheme}
\subsection{Variable Flavour Number Schemes}
\begin{figure}
\begin{center}
\scalebox{0.9}{\includegraphics{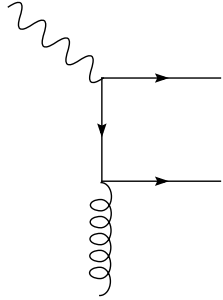}}
\caption{${\cal O}(\alpha_S)$ contribution to the boson gluon fusion process \cite{Witten}, whereby a heavy quark pair is produced in the DIS final state.}
\label{heavy}
\end{center}
\end{figure}

Heavy quark pairs in DIS can be produced by boson gluon fusion \cite{Witten} - the leading order diagram at the parton level is shown in figure \ref{heavy} \footnote{Here we are concerned with heavy quarks which are generated perturbatively, as opposed to {\it intrinsic} heavy flavour distributions, which would be present in the proton at all values of $Q^2$ \cite{Brodsky}.}. 
Higher order contributions can involve the gluon interacting further with quarks from the proton, as well as real and virtual emissions. Taking into account all such diagrams, the heavy flavour contribution to the proton structure functions $\{F_a\}$ can be written as:
\begin{equation}
F_{a,H}=\frac{1}{n_f}\sum_{i=1}^{n_f}e_i^2\left[(C_{a,qq}^{NS}+C_{a,qq}^{PS})\otimes\Sigma + C_{a,qg}^S\otimes g + n_fC_{a,qq}^{NS}\otimes q_{ns}\right]+e_H^2\left[C_{a,Hq}^{PS}\otimes\Sigma + C_{a,Hg}^S\otimes g\right],\\
\label{heavyfa}
\end{equation}
where $n_f$ is the number of active light flavours ($n_f=3$ in phenomenological applications). Following \cite{Buza2}, we have explicitly decomposed all coefficients into singlet and non-singlet parts, and also separated the coefficients into parts arising from the photon coupling to a light quark pair ($C_{a,qi}(x,Q^2/M^2)$) from those where it couples to the heavy quark pair \footnote{There are technical complications arising at NNLO due to heavy quark pairs produced away from the photon vertex, which contribute to both the light and heavy structure functions \cite{Chuvakin}. These do not concern us in this paper.} ($C_{a,Hi}(x,Q^2/M^2)$), $i\in\{q,g\}$. The coefficient $C_{a,Hq}$ has only a pure singlet contribution, as only an intermediate gluon can couple the heavy quark pair (which interacts with the photon) to a light quark pair. This is a flavour singlet exchange. The leading order of each coefficient is given in table \ref{heavycoeffs}. \begin{table}
\begin{center}
\begin{tabular}{c|c}
Coefficient & Order\\
\hline\\
$C_{a,Hg}^S$ & ${\cal O}(\alpha_S)$\\
\\
$C_{a,Hq}^{PS}$ & ${\cal O}(\alpha_S^2)$\\
\\
$C_{a,qq}^{PS,NS,S}$ & ${\cal O}(\alpha_S^2)$\\
\\
$C_{a,qg}^{S}$ & ${\cal O}(\alpha_S^3)$
\end{tabular}
\caption{The leading order of each of the coefficients in the heavy flavour structure functions.}
\label{heavycoeffs}
\end{center}
\end{table}
One could choose to use this description of the structure functions within a particular factorisation scheme at all values of $Q^2$. This is the {\it fixed flavour} number scheme. Collecting together the coefficients of each parton density, one may tidy up equation (\ref{heavy}):
\begin{align}
F_{a,H}&=C_{a,q}^{FF}\otimes\Sigma+C_{a,g}^{FF}\otimes g+C_{a,ns}^{FF}\otimes q_{ns}\\
&\equiv \vec{C}^{(FF)\,\text{\textbf{T}}}\otimes\vec{f}^{(n_f)},
\label{heavyfa2}
\end{align}
defining the fixed flavour coefficients $C_{a,i}^{FF}$ and:
\begin{eqnarray}
\vec{C}^{FF}=\left(\begin{array}{c}C_{a,q}^{FF}\\C_{a,g}^{FF}\\C_{a,ns}^{FF}\end{array}\right),&\vec{f}^{(n_f)}=\left(\begin{array}{c}\Sigma\\g\\q_{ns}\end{array}\right).
\label{vecs}
\end{eqnarray}
However, a problem arises at high $Q^2$ in that the fixed flavour coefficients are logarithmically divergent in $Q^2/M^2$. For example, one has \cite{Buza} in the $\msbar$ scheme:
\begin{equation}
C_{2,g}^{FF}\stackrel{\frac{Q^2}{M^2}\rightarrow\infty}{\longrightarrow}\left(\frac{\alpha_S}{4\pi}\right)T_R\left[(4-8x+8x^2)\left(\log\frac{Q^2}{M^2}+\log(1-x)-\log x\right)-4+32x-32x^2\right]+{\cal O}(\alpha_S^2),
\label{asympt}
\end{equation}
with $T_R=1/2$ a colour factor. In principle both the partons and the coefficient functions (beyond leading order in $\alpha_S$) are dependent on the collinear factorisation scale $\mu_F^2$. We adopt the natural scale choice $\mu_F^2=Q^2$ from now on. The problem of divergent FF coefficients can be circumvented by defining a heavy quark distribution at a matching scale $Q^2=\tilde{Q}^2$ appropriate to the threshold for heavy quark production. Singularities are then absorbed into the parton distribution, and its evolution is governed by a splitting function as in the DGLAP equations for light quarks. Below the matching scale, there is no heavy quark distribution. Different choices can be made for the matching scale - we use $\tilde{Q}^2=M^2$. This ensures that up to NLO in the fixed order expansion, the heavy partons evolve from a zero initial value. This choice also reduces technical complication at higher orders, although no choice of matching scale will ensure continuity of all partons. One now has a {\it variable flavour} number scheme, as $n_f$ (the number of active quarks) increases as $Q^2$ crosses the matching scale from below. Equation (\ref{heavyfa2}) applies for $Q^2\leq M^2$, whereas for $Q^2\geq M^2$ one has:
\begin{align}
F_{a}^H&=C_{a,H}^{VF}\otimes (q_H+\bar{q}_H)+C_{a,q}^{VF, PS}\otimes\Sigma^{(n_f+1)}+C_{a,g}^{VF}\otimes g^{(n_f+1)}\notag\\
&\equiv \vec{C}^{(VF)\,\text{\textbf{T}}}\otimes\vec{f}^{(n_f+1)},
\label{favf}
\end{align}
where $q_H$ ($\bar{q}_H$) is the quark (antiquark) distribution for the heavy flavour. Such a scheme was first suggested in \cite{ACOT}, and a proof that the variable flavour number scheme is well-defined to all orders in perturbation theory was given in \cite{Collins_heavy}.\\

The general requirements a variable flavour scheme must satisfy were derived in \cite{Buza2}. The fact that the heavy flavour structure functions cannot depend upon whether one uses a fixed or variable flavour number scheme imposes a relationship between the variable and fixed flavour partons. For the heavy flavour and singlet densities:
\begin{equation}
\vec{f}^{(n_f+1)}=\left(\begin{array}{c}q_{H+}\\g^{(n_f+1)}\\\Sigma^{(n_f+1)}\end{array}\right)=\left(\begin{array}{cc}A_{Hq}&A_{Hg}\\A_{gq}&A_{gg}\\A_{qq}&A_{qg}\end{array}\right)\left(\begin{array}{c}\Sigma^{(n_f)}\\g^{(n_f)}\end{array}\right)={\mat A}\vec{f}^{(n_f)},
\label{As}
\end{equation}
where the heavy flavour matrix elements $\{A_{ij}\}$ are perturbatively calculable in a given factorisation scheme, and $q_{H+}=q_H+\bar{q}_H$. They satisfy:
\begin{equation}
A_{ij}=\delta_{ij}\delta(1-x)+{\cal O}(\alpha_S)
\label{AijLO}
\end{equation}
n.b. at leading order, the ($n_f+1$)-flavour parton distributions are the same as their $n_f$-flavour counterparts.\\
Similarly, non-singlet quark combinations are related by:
\begin{equation}
q_{i,ns}^{(n_f+1)}=A_{qq}^{NS}q_{i,ns}^{(n_f)}.
\label{Ans}
\end{equation}
As $Q^2/M^2\rightarrow\infty$, the heavy flavour mass becomes negligible in relation to the scattering scale, and so we must be able to neglect terms $\sim{\cal O}(M^2/Q^2)$. Thus, the variable flavour coefficients must tend to the appropriate massless coefficients in the asymptotic limit of high $Q^2$. Like their fixed flavour counterparts, they are only defined up to a factorisation scheme choice. Equivalently, one is free to transform the singlet parton densities according to:
\begin{equation}
\vec{f}'^{(n_f),(n_f+1)}={\mat Z}^{(n_f),(n_f+1)}\otimes\vec{f}^{(n_f),(n_f+1)},
\label{schemes}
\end{equation}
where the matrix $\bf{\textsf{Z}}$ in each case has diagonal elements $\delta(1-x)+{\cal O}(\alpha_s)$, and off-diagonal elements ${\cal O}(\alpha_S)$. There is a further constraint from the momentum sum rule:
\begin{equation}
\int_0^1 dx x [\Sigma'^{(n_f),(n_f+1)}(x)+g'^{(n_f),(n_f+1)}(x)]=\int_0^1 dx x [\Sigma^{(n_f),(n_f+1)}(x)+g^{(n_f),(n_f+1)}(x)],
\label{momsum}
\end{equation}
where $\Sigma(x)$ is the quark singlet distribution:
\begin{equation}
\Sigma(x)=\sum_{i}[q_i(x)+\bar{q}_i(x)].
\label{sigma}
\end{equation}
The sum is over all active flavours. A similar transformation to (\ref{schemes}) also holds in the non-singlet sector, but with a $1\times 1$ transformation matrix.\\

There is a further ambiguity in the VF coefficients arising from the fact that there is one more VF coefficient than there are FF coefficients. Combining equations (\ref{favf}, \ref{As}) one finds:
\begin{align}
F_{a,H}&=(C_{a,H}^{VF}\otimes A_{Hq}+C_{a,q}^{VF,PS}\otimes A_{qq}+C_{a,g}\otimes A_{gq})\otimes\Sigma^{(n_f)}\notag\\
&\qquad+(C_{a,H}^{VF}\otimes A_{Hg}+C_{a,q}^{VF,PS}\otimes A_{qg}+C_{a,g}^{VF}\otimes A_{gg})\otimes g^{(n_f)}.
\label{fah2}
\end{align}
Equating the gluon coefficients in equations (\ref{heavyfa}, \ref{fah2}):
\begin{equation}
C_{a,H}^{VF}\otimes A_{Hg}+C_{a,q}^{VF,PS}\otimes A_{qg}+C_{a,g}^{VF}\otimes A_{gg}=C_{a,g}^{FF}.
\label{cgh}
\end{equation}
For example, using the fact that $A_{gg}=\delta(1-x)$ at leading order, one obtains at ${\cal O}(\alpha_S)$:
\begin{equation}
C_{2,H}^{VF(0)}\otimes A_{Hg}^{(1)}+C_{2,g}^{VF(1)}=C_{2,g}^{FF(1)},
\label{cgh1}
\end{equation}
where in the $\msbar$ scheme:
\begin{equation}
A_{Hg}^{(1)}=T_R\left[(4-8x+8x^2)\log\frac{Q^2}{M^2}\right]
\label{ahg1msbar}
\end{equation}
is the coefficient of $\alpha_S/(4\pi)$. Comparing with equation (\ref{asympt}), one sees that the heavy flavour matrix element contains the same divergence as the fixed flavour coefficient, so that the variable flavour coefficient is finite as $Q^2\rightarrow\infty$.\\ 

Equation (\ref{cgh}) shows that although each VF coefficient must tend to an appropriate massless limit, one can choose to shift terms vanishing at high $Q^2$ from $C_{a,H}^{VF}$ into $C_{a,g}^{VF}$. This can be examined in more detail in Mellin space, defining the moments of parton distributions \footnote{Another definition of the Mellin transform is commonly used, corresponding to $\mathbb{M}_{N-1}$ in our notation.} by:
\begin{equation}
\tilde{q}(N)=\mathbb{M}_N[q(x)]\equiv\int_0^1 dx x^N q(x)
\label{Mellin}
\end{equation}
Heavy flavour coefficients contain an implicit Heaviside function $\Theta(W^2-M^2)$, where $W^2$ is the proton-photon centre of mass energy (n.b. $W^2$ must exceed the threshold for pair production). Correspondingly, there is a maximum value of $x$:
\begin{equation}
x_{max}=\left(1+\frac{4M^2}{Q^2}\right)^{-1}.
\label{xmax}
\end{equation}
We thus define moments of the heavy coefficients via the scaled variable $x'=x/x_{max}$ to obtain:
\begin{equation}
\mathbb{M}_{N}[f(x',Q^2/M^2)]=\int_0^1 dx' x'^Nf(x',Q^2/M^2).
\label{Mellin_heavy}
\end{equation}
One then has the convolution theorem:
\begin{align}
\mathbb{M}_N[f(x',Q^2/M^2)\otimes g(x,Q^2)]&=\mathbb{M}_N\left[\int_{x'}^1 f\left(\frac{x'}{z},\frac{Q^2}{M^2}\right)g(z,Q^2)dz\right]\notag\\
&=\mathbb{M}_N[f(x',Q^2/M^2)]\mathbb{M}_N[g(x)],
\label{convol}
\end{align}
where $N$ is conjugate to $x'$ on the left-hand side, but on the right-hand side is conjugate to $x'$ in the first term of the product, and conjugate to $x$ in the second term. Taking $x'$ moments of equation (\ref{cgh}) and rearranging, one obtains the $N$-space gluon coefficient:
\begin{equation}
\tilde{C}_{a,g}^{VF}=\frac{1}{\tilde{A}_{gg}}\left[\tilde{C}_{a,g}^{FF}-\tilde{C}_{a,q}^{VF,PS}\tilde{A}_{qg}-\tilde{C}_{a,H}^{VF}\tilde{A}_{Hg}\right].
\label{cagN}
\end{equation}
Substitution of the perturbative expansions for the coefficients and heavy flavour matrix elements leads to a power series in $\alpha_S$, which can be inverse transformed term by term to give the resulting $x'$-space expression. Choosing the quark coefficients then fixes the gluon coefficient according to equation (\ref{cagN}).\\

To summarise, there are two possible types of choice in specifying a variable flavour number scheme for $Q^2\geq \tilde{Q}^2$:
\begin{enumerate}
\item{Method of collinear factorisation. Moving between different choices involves an explicit transformation of the partons according to equation (\ref{schemes}). Such an ambiguity is always present in the coefficient functions, for both the light and heavy flavours.}
\item{Position of $M^2/Q^2$ terms. Choosing a different definition of e.g. the gluon coefficient does not involve a transformation of the parton densities.}
\end{enumerate}
Consider a change of definition of type (2) above, within a fixed collinear factorisation scheme, from a scheme (A) to another scheme (B). Up to ${\cal O}(\alpha_S^3)$, one may ignore the terms involving the pure singlet coefficient and solve explicitly from equation (\ref{cagN}) for the $N$-space gluon coefficient in the new scheme:
\begin{equation}
\tilde{C}_{a,g}^{(B)VF}=[\tilde{C}_{a,H}^{(A)VF}-\tilde{C}_{a,H}^{(B)VF}]\frac{\tilde{A}_{Hg}}{\tilde{A}_{gg}}+\tilde{C}_{a,g}^{(A)VF},
\label{A2B}
\end{equation}
where the bracketed term is ${\cal O}(M^2/Q^2)$, and to leading order one may set the denominator to $1$ on the right-hand side.
\subsection{A VF Number Scheme for Small $x$ Physics}
\begin{figure}
\begin{center}
\scalebox{0.7}{\includegraphics{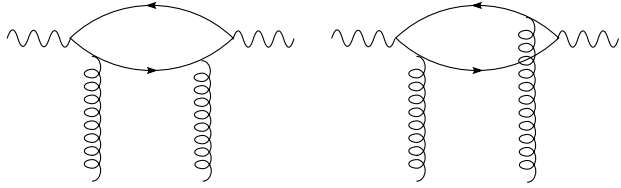}}
\caption{Diagrams contributing to the LL impact factor. Two further diagrams are obtained by reversing the direction of the quark loop.}
\label{impact}
\end{center}
\end{figure}

We now turn to the problem of how to formulate a VF scheme that can describe heavy quark production at small $x$, but also matches on to the fixed order expansion. One must first satisfy requirements (1) and (2) above, namely choose a suitable factorisation scheme and positioning of $M^2/Q^2$ terms. At small $x$ the heavy flavour contributions to the proton structure functions are given by the high energy factorisation formula \cite{Collins,CataniH}:
\begin{equation}
\tilde{F}_{a,H}=\alpha_S\int_0^\infty\frac{dk^2}{k^2}h_a(k^2/Q^2,Q^2/M^2)\tilde{f}(N,k^2,Q_0^2)\tilde{g}_B(N,Q_0^2),
\label{highfactor}
\end{equation}
where $\tilde{f}(N,k^2,Q_0^2)$ is the unintegrated gluon density, and $\tilde{g}_B(N,Q_0^2)$ a bare gluon distribution which absorbs collinear singularities upon solution of the BFKL equation. The $\{h_a\}$ are impact factors coupling the virtual photon to the gluon via a heavy quark pair (see figure \ref{impact}), whose LL forms may be found in \cite{CCH,CataniH}. The corresponding results for the light structure functions are in \cite{Catani3}. The NLL results are not yet known, but work is in progress \cite{Bartels00,Bartels01,Bartels02,Bartels04,Fadin01,Fadin02}.\\

It is tempting to immediately identify the impact factors with the heavy flavour coefficient functions $C_{2,Hg}$ and $C_{L,Hg}$, both below and above the matching point $Q^2=M^2$. However, whilst the longitudinal impact factor is finite as $Q^2/M^2\rightarrow\infty$, the factor $h_2$ diverges as a consequence of the fact that $F_2$ is proportional to the quark singlet distribution which is intrinsically non-perturbative ($F_L$, on the other hand, begins at ${\cal O}(\alpha_S)$). Instead one may consider the quantity:
\begin{equation}
\frac{\partial \tilde{F}_{2,H}}{\partial\log Q^2}=\alpha_S\int_0^\infty \frac{dk^2}{k^2}\tilde{h}_2(k^2/Q^2,Q^2/M^2) \tilde{f}(N,k^2,Q_0^2)\tilde{g}_B(N,Q_0^2),
\label{f2deriv}
\end{equation}
which serves to define the factor $\tilde{h}_2$. The resulting coefficient is then not simply $C_{2,Hg}$, but in general is a mixture of the coefficient function and anomalous dimensions $\gamma_{Hg}$, $\gamma_{qg}$ entering the DGLAP equation and governing the evolution of the charm and singlet distributions. The interpretation of this quantity depends on the factorisation scheme. In the massless case (recovered here by taking $Q^2/M^2\rightarrow\infty$), it is simplest to adopt the DIS scheme \cite{Altarelli}, in which the structure function $F_2$ has the na\"{i}ve parton model form:
\begin{equation}
F_2(x,Q^2)=\sum_{i=1}^{n_f}[e_i^2[q_i(x,Q^2)+\bar{q}_i(x,Q^2)]
\label{naivef2}
\end{equation}
to all orders, where $e_i$ is the electromagnetic charge of quark species $q_i$. The impact factor $\tilde{h}_2$ can then be interpreted as the quark-gluon anomalous dimension $\gamma^{\DIS}_{qg}$. If one were to adopt this in the massive case, however, the massive impact factor $\tilde{h}_2$ would be interpreted as an $M^2$-dependent anomalous dimension coupling the heavy flavour distribution to the gluon. This is undesirable for a number of reasons. Firstly, there is difficulty conserving momentum either side of the matching point if splitting functions of massless quarks differ from those of the heavy species. Furthermore, mass-dependent splitting functions will lend considerable analytic complication to the fixed order expansion. The massless $\msbar$ scheme anomalous dimensions are well known (now up to ${\cal O}(\alpha_S^3)$ \cite{Vogt_s,Vogt_ns}), and can be straightforwardly transformed to other schemes. More importantly, the use of massive splitting functions for the heavy partons makes them different to the massless partons.  Consider the structure functions at high $Q^2$. The fact that the massive partons have evolved differently to the massless ones from their matching scales up to high $Q^2$ means that the coefficient functions are also different, and do not reduce to the conventional massless coefficients \cite{Olness}. Such a division between the parton species seems unphysical.\\

Instead it makes sense to factorise the partons in the DIS scheme such that both the heavy and the light flavours evolve according to the massless DIS splitting functions. One can ensure this by transforming from the $\msbar$ scheme partons. The explicit $N$-space transformation for $Q^2\leq M^2$ is entirely mass-independent and takes the form (see e.g. \cite{Vogts}):
\begin{equation}
\left(\begin{array}{c}\tilde{\Sigma}^{\DIS(n_f)}\\\tilde{g}^{\DIS(n_f)}\end{array}\right)=\left(\begin{array}{cc}\tilde{C}_{2,q}^{\msbar}&2n_f\tilde{C}_{2,g}^{\msbar}\\1-\tilde{C}_{2,q}^{\msbar}&1-2n_f\tilde{C}_{2,g}^{\msbar}\end{array}\right)\left(\begin{array}{c}\tilde{\Sigma}^{\msbar(n_f)}\\\tilde{g}^{\msbar(n_f)}\end{array}\right),
\label{ms2dis}
\end{equation}
where flavour factors have been explicitly displayed in the massless $\msbar$ scheme coefficients. This leads to a similar transformation matrix for the singlet and heavy quark distributions above the matching point\footnote{Strictly speaking there is a further pure singlet contribution involved in the transformation of the single quark species $q_{H+}$. This can be neglected up to ${\cal O}(\alpha_S^2)$.}:
\begin{equation}
\left(\begin{array}{c}\tilde{q}_{H+}^\DIS\\\tilde{g}^\DIS\\\tilde{\Sigma}^\DIS\end{array}\right)=\left(\begin{array}{ccc}\tilde{C}_{2,q}^{\msbar}&2\tilde{C}_{2,g}^{\msbar}&0\\1-\tilde{C}_{2,q}^{\msbar}&1-2(n_f+1)\tilde{C}_{2,g}^{\msbar}&1-\tilde{C}_{2,q}^{\msbar}\\0&2(n_f+1)\tilde{C}_{2,g}^{\msbar}&\tilde{C}_{2,q}^{\msbar}\end{array}\right)\left(\begin{array}{c}\tilde{q}_{H+}^{\msbar}\\\tilde{g}^{\msbar}\\ \Sigma^{\msbar}\end{array}\right),
\label{ms2dish}
\end{equation}
with $q_{H+}=q_H+\bar{q}_H$. For the non-singlet sector:
\begin{equation}
\tilde{q}_{ns}^{\DIS(n_f)}=\tilde{C}_{ns}^{\msbar}\tilde{q}_{ns}^{\msbar(n_f)}.
\label{transns}
\end{equation}
The DIS scheme partons now have no mass dependence, other than that introduced by the transition points $Q^2=M^2$ for each heavy flavour. Instead, explicit dependence on $M^2/Q^2$ resides in the FF and VF coefficient functions, rather than the anomalous dimensions governing the parton evolution. The transformations above explicitly define a collinear factorisation scheme, thus fulfilling condition (1) above.\\

One must now specify the position of ${\cal O}(M^2/Q^2)$ terms. It has been suggested previously in the $\msbar$ scheme fixed order expansion \cite{ACOTchi} that one should choose the leading order contribution to the $F_2^H$ heavy quark coefficient to be $C_{2,H}^{\msbar(0)VF}=\delta(1-x')$. This is known as the ACOT($\chi$) scheme. Note that this automatically generates the correct asymptotic limit at high $Q^2$, where $x'\rightarrow x$. One can extend this to all orders i.e. $C_{2,H}^{\msbar,VF}=C_{2,q}^{\msbar}(x')$ to give a simple, physically motivated definition of the quark coefficient at all orders. A similar definition can also be chosen for the heavy quark coefficient entering $F_{L,H}$ \cite{Thorne06}.\\

We propose using the same specification for the DIS scheme. Here, this amounts to choosing $C_{2,H}^{\DIS, VF}=\delta(1-x')$ to all orders in $\alpha_S$. It will be seen that this simplifies considerably the solution of the gluon coefficient in the small $x$ limit. This is clearly a physical way of defining the heavy quark coefficient within the DIS scheme, which is itself a more physical scheme to work in due to its connection with the na\"{i}ve parton model. The natural choice for $F_{L,H}$ is also to choose $C_{L,H}^{\DIS, VF}=C_{L,q}^{\DIS}(x')$, where the right-hand side denotes the functional form of the massless coefficient. However, a problem occurs with smoothness of $F_{L,H}$ across the matching point. At ${\cal O}(\alpha_S)$ in the longitudinal case, equation (\ref{cgh}) gives:
\begin{equation}
C_{L,g}^{VF(1)}=C_{L,g}^{FF(1)},
\label{clgvf-ff}
\end{equation}
and one has \cite{Witten}:
\begin{equation}
C_{L,g}^{FF(1)}=4x(1-x)\left[2v-(1-v^2)\ln\left(\frac{1+v}{1-v}\right)\right]\Theta(W^2-4M^2),
\label{clgff1exp}
\end{equation}
with:
\begin{equation}
v^2=1-\frac{4M^2x}{Q^2(1-x)}.
\label{v}
\end{equation}
The parameter $v$ is the velocity of the heavy quark or antiquark in the photon-proton centre of mass frame, and hence satisfies $0\leq v< 1$. Taylor expanding (\ref{clgff1exp}) gives:
\begin{equation}
C_{L,g}^{FF(1)}=8x(1-x)v^3+{\cal O}(v^4).
\label{clgff1v}
\end{equation}
The fixed and variable flavour descriptions of $F_{L,H}$ dictate at ${\cal O}(\alpha_S)$:
\begin{align}
F_{L,H}^{FF}&=\left(\frac{\alpha_S}{4\pi}\right)C_{L,g}^{FF(1)}\otimes g^{(n_f)}+{\cal O}(\alpha_S^2);\label{FLH1}\\
F_{L,H}^{VF}&=\left(\frac{\alpha_S}{4\pi}\right)\left[C_{L,g}^{VF(1)}\otimes g^{(n_f+1)}+C_{L,H}^{VF(1)}\otimes q_{H+}\right]+{\cal O}(\alpha_S^2).
\label{FLH2}
\end{align}
The FF contribution and the first term of equation (\ref{FLH2}) are suppressed by $v^3$, whereas the second term in equation (\ref{FLH2}) is not if one chooses $C_{L,H}^{VF}=C_{L,q}^{\DIS}(x')$. Even though the heavy quark distribution will be small at the matching scale, one will still have a kink in the structure function at $Q^2=M^2$. Smoothness of the structure function can be recovered by instead demanding $C_{L,H}^{VF}=f(Q^2/M^2)C_{L,q}^{\DIS}(x')$, where $f(Q^2/M^2)$ is a function such that $f(1)=0$ and $f(\infty)=1$ with a smooth interpolation between these values. A suitable choice is \cite{Thorne06}:
\begin{equation}
f\left(\frac{Q^2}{M^2}\right)=\frac{5}{4}\left(\frac{1}{1+4M^2/Q^2}-\frac{1}{5}\right),
\label{f}
\end{equation}
which fulfils the above conditions and is suppressed by the heavy mass factor $4M^2$ at intermediate $Q^2$ values. Note that $C_{L,g}^{VF(1)}$ is not affected by this choice, but is the same as that obtained by choosing $f=1$. This follows from equation (\ref{A2B}), which at ${\cal O}(\alpha_S)$ gives $C_{L,g}^{(A)(1)}=C_{L,g}^{(B)(1)}$ when transforming between schemes differing only in the placement of ${\cal O}(M^2/Q^2)$ terms. The NLO coefficient $C_{Lg}^{VF(2)}$ and higher orders, however, will be affected as they depend upon the choice of $C_{L,H}^{VF(1)}$. Note the prefactor of equation (\ref{f}) is chosen to have no dependence on $x$, thus introduces no extra complication in convolutions involving the heavy quark coefficient, matrix elements and the partons. \\

There is a further problem in $F_{2,H}$, regarding the order of the perturbation expansion either side of the matching scale. For $Q^2\leq M^2$, one has $F_{2,H}\sim{\cal O}(\alpha_S)$, whereas $F_{2,H}\sim{\cal O}(\alpha_S^0)$ for $Q^2>M^2$. One must choose how to deal with the ordering in such as way as to guarantee continuity of $F_{2,H}$ in the fixed order expansion. Following \cite{RT}, we add the frozen contribution:
\begin{equation}
F_{2,H}^{VF(0)}=C_{2,Hg}^{FF}(x,Q^2=M^2)\otimes g(x,Q^2=M^2)
\label{frozen}
\end{equation}
for $Q^2>M^2$ (see \cite{Thorne06} for a discussion).\\

One has now satisfied choice (2) above for both $F_{2,H}$ and $F_{L,H}$, and thus completely defined a variable flavour number scheme. By analogy with ACOT's proposal, we refer to this as the DIS($\chi$) scheme. In the next section, we will present the derivation of the small $x$ variable flavour (and fixed flavour) gluon coefficient functions. To use them alongside the fixed order expansion, however, the known $\msbar$ scheme heavy flavour coefficient functions and matrix elements must be transformed into the $\DIS$ scheme. \\

Transformation of the FF coefficients proceeds from equations (\ref{heavyfa2}, \ref{ms2dis}, \ref{transns}), together with scheme independence of the structure functions:
\begin{equation}
F_{a,H}=\vec{C}^{(FF,\DIS)\,\text{\textbf{T}}}\otimes \vec{f}^{(n_f)\DIS}=\vec{C}^{(FF,\msbar)\,\text{\textbf{T}}}\otimes \vec{f}^{(n_f)\msbar}.
\label{schemeindep}
\end{equation}
One obtains:
\begin{equation}
(\begin{array}{ccc}C_{a,q}^{FF,\DIS}&C_{a,g}^{FF,\DIS}&C_{a,ns}^{FF,\DIS}\end{array})=(\begin{array}{ccc}C_{a,q}^{FF,\msbar}&C_{a,g}^{FF,\msbar}&C_{a,ns}^{FF,\msbar}\end{array})\otimes\left(\begin{array}{cc}{\mat Z}^{(n_f)}&0\\0&C_{2,ns}^{\msbar}\end{array}\right)^{-1},
\label{transFF}
\end{equation}
where ${\mat Z}^{(n_f)}$ is the matrix of equation (\ref{ms2dis}). Here inversion of the matrix under a convolution sign can be understood by transforming to $N$-space, substituting in the perturbative expansions of all quantities, then inverse transforming term by term back to $x'$-space. A similar procedure can be carried out for the variable flavour coefficient functions, and one obtains:
\begin{equation}
(\begin{array}{ccc}C_{a,H}^{VF,\DIS}&C_{a,g}^{VF,\DIS}&C_{a,q}^{VF,\DIS}\end{array})=(\begin{array}{ccc}C_{a,H}^{VF,\msbar}&C_{a,g}^{VF,\msbar}&C_{a,q}^{VF,\msbar}\end{array})[{\mat Z}^{(n_f+1)}]^{-1},
\label{transVF}
\end{equation}
with ${\mat Z}^{(n_f+1)}$ the matrix of equation (\ref{ms2dish}). Explicit results up to ${\cal O}(\alpha_S)$ for the FF coefficients are:
\begin{align}
C_{2,g}^{FF,\DIS(1)}&=C_{2,g}^{FF,\msbar(1)};\label{c2gff1}\\
C_{L,g}^{FF,\DIS(1)}&=C_{L,g}^{FF,\msbar(1)}.\label{clgff1}
\end{align}
For the VF coefficients:
\begin{align}
C_{2,H}^{VF,\DIS(0)}&=C_{2,H}^{VF,\msbar(0)}=\delta(1-x');\label{c2hvf0}\\
C_{2,H}^{VF,\DIS(1)}&=0;\label{c2hvf1}\\
C_{2,g}^{VF,\DIS(1)}&=C_{2,g}^{VF,\msbar(1)}-2C_{2,g}^{\msbar(1)};\label{c2gvf1}\\
C_{L,H}^{VF,\DIS(1)}&=C_{L,H}^{VF,\msbar(1)};\label{clhvf1}\\
C_{L,g}^{VF,\DIS(1)}&=C_{L,g}^{VF,\DIS(1)}.\label{clgvf1}
\end{align}
Note that to this order one has $C_{2,H}^{VF,\DIS}=C_{2,H}^{\DIS}(x')$ as required by the choice of where to put ${\cal O}(M^2/Q^2)$ terms, thus there is no need to perform a further transformation according to equation (\ref{A2B}).\\

One must also transform the heavy flavour matrix elements $\{A_{ij}\}$ from the $\msbar$ scheme, where they are known up to ${\cal O}(\alpha_S^2)$ \cite{Buza2}, to the DIS scheme. This can be done using scheme independence of $F_2$, as well as equivalence of the FF and VF descriptions:
\begin{align*}
F_{2,H}&=\vec{C}^{(VF,\DIS)\,\text{\textbf{T}}}\otimes\vec{f}^{{(n_f+1)\DIS}}\\
&=\vec{C}^{(VF,\DIS)\,\text{\textbf{T}}}\otimes{\mat A}^{\DIS}\otimes\vec{f}^{(n_f)\DIS}\\
&=\vec{C}^{(VF,\msbar)\,\text{\textbf{T}}}\otimes[{\mat Z}^{(n_f+1)}]^{-1}\otimes{\mat A}^{\DIS}\otimes{\mat Z}^{(n_f)}\otimes\vec{f}^{(n_f)\msbar}\\
&=\vec{C}^{(VF,\msbar)\,\text{\textbf{T}}}\otimes{\mat A}^{\msbar}\otimes\vec{f}^{{(n_f)\msbar}}
\end{align*}
and therefore:
\begin{equation}
{\mat A}^{\DIS}={\mat Z}^{(n_f+1)}\otimes{\mat A}^{\msbar}\otimes[{\mat Z}^{(n_f)}]^{-1}.
\label{transAij}
\end{equation}
Explicit results up to ${\cal O}(\alpha_S)$ are:
\begin{align}
A_{gg}^{\DIS(0)}&=A_{gg}^{\msbar(0)}=\delta(1-x');\label{agg0}\\
A_{gg}^{\DIS(1)}&=A_{gg}^{\msbar(1)}-2C_{2g}^{\msbar(1)};\label{agg1}\\
A_{Hg}^{\DIS(1)}&=A_{Hg}^{\msbar(1)}+2C_{2g}^{\msbar(1)}\label{ahg1}.
\end{align}
Expanding equation (\ref{cgh}) to ${\cal O}(\alpha_S)$ in the $\msbar$ scheme, one finds:
\begin{equation}
C_{2,g}^{FF,\msbar(1)}=C_{2,g}^{VF,\msbar(1)}-A_{Hg}^{\msbar(1)}.
\label{cgvf}
\end{equation}
Applying the results of equations (\ref{c2gvf1}, \ref{c2gff1}, \ref{ahg1}) one then finds:
\begin{equation}
C_{2,g}^{FF,\DIS(1)}=C_{2,g}^{VF,\DIS(1)}-A_{Hg}^{\DIS(1)},
\label{cgvf2}
\end{equation}
which is exactly what one expects in the DIS scheme, given that in equation (\ref{cgh}) no factorisation scheme is specified. This serves as a consistency check of the above transformations.\\
\subsection{The $\DIS(\chi)$ Scheme at Small $x$}
To see how the above scheme is implemented in the small $x$ limit, it is convenient to Mellin transform equation (\ref{highfactor}) according to:
\begin{equation}
\tilde{F}(\gamma,N)=\int_0^\infty dk^2 (k^2)^{-1-\gamma}F(k^2,N).
\label{Mellin2}
\end{equation}
We now drop tildes in $N$ and $\gamma$ space, instead denoting explicitly the arguments of each function. For $F_{2,H}$:
\begin{align}
F_{2,H}(\gamma,N,Q^2/M^2)&=h_2(\gamma,N,Q^2/M^2)f(N,\gamma,Q_0^2)g_B(Q_0^2,N)\notag\\
&=h_2(\gamma,N,Q^2/M^2)g(N,\gamma).
\label{f2dm}
\end{align}
For $Q^2\leq M^2$, one identifies:
\begin{equation}
C_{2,g}^{FF}(\gamma,N,Q^2/M^2)=h_2(\gamma,N,Q^2/M^2).
\label{c2gff}
\end{equation}
For $Q^2>M^2$, one must consider the VF expression for $F_2$ in the $\DIS(\chi)$ scheme. As has already been noted, the impact factor $h_2$ diverges as $Q^2/M^2\rightarrow\infty$  due to lack of collinear safety, thus one cannot directly take the variable flavour coefficient from $h_2$. However, one can make progress from equation (\ref{fah2}) expressing the relationship between the VF and FF schemes. Ignoring the pure singlet term in equation (\ref{cgh}), and using $A_{gg}=\delta(1-x)$ and $C_{2,H}^{VF}=\delta(1-x')$ at LL order (where the second equality follows from the definition of the $\DIS(\chi)$ scheme):
\begin{equation}
A_{Hg}(\gamma,N,Q^2/M^2)=C_{2,g}^{FF}(\gamma,N,Q^2/M^2)-C_{2,g}^{VF}(\gamma,N,Q^2/M^2)
\label{ahg}
\end{equation}
Given that the equality holds for any value of $Q^2$, one may consider the limit $Q^2\rightarrow\infty$. One must have:
\begin{equation}
\lim_{\frac{Q^2}{M^2}\rightarrow\infty}C_{2,g}^{VF}(\gamma,N,Q^2/M^2)=0,
\label{boundary}
\end{equation}
i.e. the variable flavour gluon coefficient tends to the appropriate massless limit at high $Q^2$, which is zero. Then the second term in equation (\ref{ahg}) vanishes and leaves:
\begin{equation}
A_{Hg}(\gamma,N,Q^2/M^2)=C_{2,g}^{FF}(\gamma,N,Q^2/M^2)|_{\frac{Q^2}{M^2}\rightarrow\infty},
\label{ahg2}
\end{equation}
where the notation on the right-hand side implies that the parts which vanish as $Q^2/M^2\rightarrow\infty$ are removed. Thus, the $\DIS(\chi)$ VF gluon coefficient at small $x$ is given in double Mellin space by:
\begin{align}
C_{2,g}^{VF}(\gamma,N,Q^2/M^2)&=C_{2,g}^{FF}(\gamma,N,Q^2/M^2)-A_{Hg}(\gamma,N,Q^2/M^2)\notag\\
&=h_2(\gamma,N,Q^2/M^2)-h_2(\gamma,N,Q^2/M^2)|_{\frac{Q^2}{M^2}\rightarrow\infty}
\label{c2gvf}
\end{align}
It is not immediately obvious that all of the requirements of the $\DIS(\chi)$ scheme have been satisfied. It must be checked that the partons evolve according to the massless DIS scheme splitting functions, and that the heavy quark coefficient is trivially defined to all orders. One way to show that these conditions are implemented successfully is to derive the above result in a different way, which highlights their role explicitly. \\

Considering the derivative of $F_{2,H}$ with respect to $\log{Q^2}$, one has:
\begin{align}
\frac{\partial F_{2,H}(\gamma,N,Q^2/M^2)}{\partial\log{Q^2}}&=\left(\frac{\partial C_{2,g}^{VF}(\gamma,N,Q^2/M^2)}{\partial\log{Q^2}}+C_{2,g}^{VF}(\gamma,N,Q^2/M^2)\gamma+C_{2,H}^{VF}(\gamma,N,Q^2/M^2)\gamma_{qg}(\gamma,N,Q^2/M^2)\notag\right.\\
&\left.\quad +2(n_f+1)C_{2,q}^{PS,VF}(\gamma,N,Q^2/M^2)\gamma_{qg}(\gamma,N,Q^2/M^2)\right)g(\gamma,N)+\ldots,\notag\\
&=\tilde{h}_{2}(\gamma,N,Q^2/M^2)g(\gamma,N),
\label{h2h}
\end{align}
where the ellipsis denotes terms in partons other than the gluon. The final term in the bracket can be neglected to LL order, as at lowest order in the small $x$ expansion $\gamma_{qg}$ and $C_{2,H}^{PS,VF}$ both contain NLL terms. In $N$-space, these correspond to contributions of the form $\sim \alpha_S(\alpha_S/N)^n$. The product of two such terms is:
\begin{equation}
\alpha_S\left(\frac{\alpha_S}{N}\right)^n\alpha_S\left(\frac{\alpha_S}{N}\right)^m=\alpha_S^2\left(\frac{\alpha_S}{N}\right)^{n+m},
\label{NNLL}
\end{equation}
which gives a NNLL contribution. The impact factor $\tilde{h}_2$ is related to $h_2$ by:
\begin{equation}
\tilde{h}_2(\gamma, N, Q^2/M^2)=\gamma h_2(\gamma,N,Q^2/M^2)+\frac{\partial h_2(\gamma,N,Q^2/M^2)}{\partial\log{Q^2}}.
\label{h2m}
\end{equation}
One now implements the $\DIS(\chi)$ scheme by identifying $\gamma_{qg}(\gamma,N,Q^2/M^2)$ with the limit of $\tilde{h}_2(\gamma,N,Q^2/M^2)$ as $Q^2/M^2\rightarrow\infty$ (this corresponds to the massless DIS scheme anomalous dimension $\gamma_{qg}^{\DIS}(\gamma,N)$), and setting $C_{2,H}^{VF}(\gamma,N,Q^2/M^2)=1$. Rearranging equation (\ref{h2h}) then gives:
\begin{align}
\frac{\partial C_{2,g}^{VF}(\gamma,N,Q^2/M^2)}{\partial\log{Q^2}}+\gamma C_{2,g}^{VF}(\gamma,N,Q^2/M^2)&=\frac{\partial h_2(\gamma,N,Q^2/M^2)}{\partial\log{Q^2}}-\left.\frac{\partial h_2(\gamma,N,Q^2/M^2)}{\partial\log{Q^2}}\right|_{\frac{Q^2}{M^2}\rightarrow\infty}\notag\\
&+\gamma \left[h_2(\gamma,N,Q^2/M^2)-h_2(\gamma,N,Q^2/M^2)|_{\frac{Q^2}{M^2}\rightarrow\infty}\right],
\label{diffeq2}
\end{align}
to be solved subject to the boundary condition of equation (\ref{boundary}). The solution is then given by equation (\ref{c2gvf}), as found previously by considering the equivalence of the FF and VF descriptions for $F_2$. The former method explicitly demonstrates continuity of the structure function at small $x$ across the matching scale $Q^2=M^2$. The second method involving the derivative of $F_2$ allows one to see how the definition of $C_{2,g}^{VF}$ relies upon the $\DIS(\chi)$ scheme choice of splitting functions and heavy quark coefficient. \\

For the heavy flavour contribution to the longitudinal structure function, one has in double Mellin space for $Q^2\leq M^2$:
\begin{equation}
F_{L,H}=h_{L}(\gamma, N, Q^2/M^2)g(\gamma,N).
\label{flkt}
\end{equation}
For $Q^2>M^2$, implementation of the $\DIS(\chi)$ scheme is somewhat simpler than in the case of $F_{2,H}$. The longitudinal impact factor is not divergent as $Q^2/M^2\rightarrow\infty$, and so one can apply the factorisation formula (\ref{flkt}) at all scales. Thus the coefficient $C_{L,g}^{VF}(\gamma,N,Q^2/M^2)$ at small $x$ is obtained from the impact factor $h_{L}(\gamma,N,Q^2/M^2)$ as for $C_{L,g}^{FF}$.\\

There is potentially a discontinuity in $F_{L,H}$ across the matching point in the small $x$ limit, arising from the non-zero initial value of the heavy parton distribution. Analogously to equations (\ref{FLH1},\ref{FLH2}) one has:
\begin{align}
F_{L,H}^{FF}(x',\alpha_S(Q^2),Q^2/M^2)&=C_{L,g}^{FF}(x',Q^2/M^2)\otimes g(x,Q^2);\label{FLH1rs}\\
F_{L,H}^{VF}(x',\alpha_S(Q^2),Q^2/M^2)&=C_{L,H}^{VF}(x',Q^2/M^2)\otimes q_{H+}+C_{L,g}^{VF}(x',Q^2/M^2)\otimes g(x,Q^2)\notag\\
&\equiv C_{L,H}^{VF}(x',Q^2/M^2)\otimes A_{Hg}(x',Q^2/M^2)\otimes g(x,Q^2)+C_{L,g}^{VF}(x',Q^2/M^2)\otimes g(x,Q^2).\label{FLH2rs}
\end{align}
At $Q^2=M^2$, the FF contribution and the second term in equation (\ref{FLH2rs}) are equal, as the FF and VF gluon coefficients are both obtained from the same impact factor. The potential discontinuity arises from the first term of equation (\ref{FLH2rs}), although this is formally ${\cal O}(\alpha_S^2(\alpha_S/N)^n)$ in the high energy expansion given that $A_{Hg}$ and $C_{L,H}$ both contain terms in $\alpha_S(\alpha_S/N)$ as their leading high energy divergence \footnote{A similar problem occurs in the fixed order expansion, where there is a discontinuity at ${\cal O}(\alpha_S^3)$, due to the non-zero value of the NNLO matrix element $A_{Hg}^{(2)}$ at $Q^2=M^2$ (n.b. at ${\cal O}(\alpha_S^2)$). See \cite{Thorne06} for a discussion.}. However, the prefactor $f(Q^2/M^2)$ of equation (\ref{f}) suppresses the first term of equation (\ref{FLH2rs}), which is then zero at $Q^2=M^2$ itself, thus facilitating smoothness of the structure function. Nevertheless, continuity of $F_{L,H}$ is only formally restored by taking higher orders in the resummed coefficient.\\

Care must be taken in implementing the resummed VF coefficient, due to the frozen term at $Q^2=M^2$ used in the fixed order expansion (equation (\ref{frozen})). The VF coefficient implicitly contains the fixed flavour coefficient according to equation (\ref{c2gvf}). Thus, one must subtract the ${\cal O}(\alpha_S)$ contribution to the resummed $C_{2,g}^{FF}$ from the resummed VF coefficient to avoid counting this term twice at $Q^2=M^2$, as it is already included in the fixed order expansion. Explicitly, the FF description at LO and LL order gives:
\begin{equation}
F_{2,H}^{FF}=\frac{\alpha_S}{4\pi}C_{2,g}^{FF(1)}\left(x',\frac{Q^2}{M^2}\right)\otimes g^{(n_f)}(x,Q^2)+\bar{C}_{2,g}^{FF,rs}\left(x',\frac{Q^2}{M^2}\right)\otimes g^{(n_f)}(x,Q^2),
\label{c2gffsplit}
\end{equation}
where:
\begin{equation}
\bar{C}_{2,g}^{FF,rs}\left(x',\frac{Q^2}{M^2}\right)=C_{2,g}^{FF,rs}\left(x',\frac{Q^2}{M^2}\right)-\frac{\alpha_S}{4\pi}C_{2,g}^{FF,rs(1)}\left(x',\frac{Q^2}{M^2}\right)
\label{c2gffbar}
\end{equation}
is the total resummed FF coefficient with its leading term subtracted (n.b. this remainder is ${\cal O}(\alpha_S^2)$), and $C_{2,g}^{FF,rs(1)}$ denotes the ${\cal O}(\alpha_S)$ contribution to the resummed coefficient. One subtracts the ${\cal O}(\alpha_S)$ term from the resummation, as this is already included in $C_{2,g}^{FF(1)}$. Na\"{i}vely in the VF description, one would write:
\begin{equation}
F_{2,H}^{VF}=q_{H+}+\frac{\alpha_S}{4\pi}\left[C_{2,g}^{FF(1)}(x',1)+C_{2,g}^{VF,rs(1)}\left(x',\frac{Q^2}{M^2}\right)\right]\otimes g^{(n_f+1)}+\bar{C}_{2,g}^{VF,rs}\otimes g^{(n_f+1)},
\label{c2gvfsplit}
\end{equation}
where the frozen term is included in the fixed order expansion at ${\cal O}(\alpha_S)$. At LO and LL order, the ${(n_f+1)}$-flavour partons are the same as their $n_f$-flavour counterparts, and using equation (\ref{boundary}) in $x$-space, one obtains:
\begin{align}
F_{2,H}^{VF}&=\left\{\frac{\alpha_S}{4\pi}A_{H,g}^{rs(1)}\left(x',Q^2/M^2\right)+\bar{A}_{Hg}^{rs}\left(x',\frac{Q^2}{M^2}\right)\notag\right.\\
&+\frac{\alpha_S}{4\pi}\left[C_{2,g}^{FF(1)}(x',1)+C_{2,g}^{FF,rs(1)}\left(x',\frac{Q^2}{M^2}\right)-A_{Hg}^{rs(1)}\left(x',\frac{Q^2}{M^2}\right)\right]\notag\\
&\left.+\bar{C}_{2,g}^{FF,rs}\left(x',\frac{Q^2}{M^2}\right)-\bar{A}_{H,g}^{rs}\left(x',\frac{Q^2}{M^2}\right)\right\}\otimes g^{(n_f)}(x,Q^2),
\label{c2gvfsplit2}
\end{align}
where the first line originates from the heavy quark. Evaluating this at $Q^2=M^2$, one obtains:
\begin{equation}
F_{2,H}^{VF}=\left\{\frac{\alpha_S}{4\pi}\left[C_{2,g}^{FF(1)}(x',1)+C_{2,g}^{FF,rs(1)}(x',1)\right]+\bar{C}_{2,g}^{FF,rs}(x',1)\right\}\otimes g(x,M^2).
\label{c2gvfsplit3}
\end{equation}
Comparing this with equation (\ref{c2gffsplit}), one sees there is a double-counting of the high energy limit of the LO FF coefficient, present in both the frozen term and the VF resummation. One must therefore subtract $C_{2,g}^{FF,rs(1)}(x',1)$ (where $x'$ is evaluated at $Q^2=M^2$) from the resummed VF coefficient. Equation (\ref{c2gvfsplit}) then becomes:
\begin{align}
F_{2,H}^{VF}&=q_{H+}+\frac{\alpha_S}{4\pi}\left[C_{2,g}^{FF(1)}(x',1)+C_{2,g}^{VF,rs(1)}\left(x',\frac{Q^2}{M^2}\right)-C_{2,g}^{FF,rs(1)}(x',1)\right]\otimes g^{(n_f+1)}\notag\\
&\quad+\bar{C}_{2,g}^{VF,rs}\otimes g^{(n_f+1)}.
\label{c2gvfsplit4}
\end{align}
\subsection{The Heavy Flavour Matrix Element $A_{Hg}$}
A further check of the above results is obtained by considering the LL limit of $A_{Hg}$, as given by equation (\ref{ahg2}). Using the form of the impact factor $h_2(\gamma,N)$ \cite{CCH}, one finds:
\begin{align}
A_{Hg}(\gamma,N,Q^2/M^2)&=h_2(\gamma,N,Q^2/M^2)|_{\frac{Q^2}{M^2}\rightarrow\infty}\notag\\
&=\frac{3h(\gamma)}{16\pi^2}\frac{4^{1-\gamma}}{(7-5\gamma)(1+2\gamma)}\left\{\frac{2\gamma^2-\gamma-1}{\gamma}\left(\frac{Q^2}{4M^2}\right)^{-\gamma}+\frac{\sqrt{\pi}(2+3\gamma-3\gamma^2)}{2}\frac{\Gamma(\gamma)}{\Gamma(1/2+\gamma)}\right\},
\label{limh2}
\end{align}
with:
\begin{equation}
h(\gamma)=\frac{4\pi}{3}\alpha_S \frac{7-5\gamma}{3-2\gamma}\frac{\Gamma^3(1-\gamma)\Gamma(1+\gamma)}{\Gamma(2-2\gamma)}.
\label{h}
\end{equation}
Expanding equation (\ref{limh2}) in $\gamma$ yields:
\begin{equation}
A_{Hg}(\gamma,N,Q^2/M^2)=\frac{\alpha_S}{4\pi}\left\{\left(\frac{4}{3}\log\frac{Q^2}{M^2}+\frac{2}{3}\right)+\gamma\left(-\frac{2}{3}\log^2\frac{Q^2}{M^2}+\frac{20}{9}\log\frac{Q^2}{M^2}+\frac{10}{9}-\frac{2\pi^2}{9}\right)+{\cal O}(\gamma^2)\right\}.
\label{ahg3}
\end{equation}
At LL order with fixed coupling $\alpha_S$, $\gamma$ is identified with the BFKL anomalous dimension \cite{BFKL}:
\begin{equation}
\gamma_{\text{BFKL}}=\frac{\bar{\alpha}_S}{N}+2\zeta(3)\left(\frac{\bar{\alpha}_S}{N}\right)^4+2\zeta(5)\left(\frac{\bar{\alpha}_S}{N}\right)^6+{\cal O}\left[\left(\frac{\bar{\alpha}_S}{N}\right)^7\right]
\label{gamBFKL}
\end{equation}
and so the high energy N-space heavy flavour matrix element is given by:
\begin{equation}
A_{Hg}(N,Q^2/\Lambda^2,Q^2/M^2)=\frac{\alpha_S}{4\pi}\left[\frac{4}{3}\log\frac{Q^2}{M^2}+\frac{2}{3}\right]+\left(\frac{\alpha_S}{4\pi}\right)^2\left[-8\log^2\frac{Q^2}{M^2}+\frac{80}{3}\log\frac{Q^2}{M^2}+\frac{40}{3}-\frac{8}{3}\pi^2\right]\frac{1}{N}+{\cal O}(\alpha_S^3).
\label{ahg4}
\end{equation}
This can be compared with the $N\rightarrow0$ limit of the fixed order matrix element in the $\DIS$ scheme. For the ${\cal O}(\alpha_S^2)$ contribution, one needs the result (derived in $x$-space from equation (\ref{transAij})):
\begin{align}
A_{Hg}^{\DIS(2)}&=A_{Hg}^{\msbar(2)}+2C_{2,g}^{\msbar(2)}+2C_{2,g}^{\msbar(1)}\otimes A_{gg}^{\msbar(1)}+2n_fC_{2,g}^{\msbar(1)}\otimes A_{Hg}^{\msbar(1)}\notag\\
&\qquad+4n_fC_{2,g}^{\msbar(1)}\otimes C_{2,g}^{\msbar(1)}+C_{2,q}^{\msbar(1)}\otimes A_{Hg}^{\msbar(1)}.
\label{transahg2}
\end{align}
The corresponding result for $A_{Hg}^{\DIS(1)}$ is given in equation (\ref{ahg1}). At ${\cal O}(\alpha_S^2)$, $A_{Hg}^{\msbar}$ can be found in \cite{Buza2}. Applying the above transformations yields:
\begin{align}
A_{Hg}^{\DIS(1)}\left(x,\frac{Q^2}{M^2}\right)&=2\left[4[x^2+(1-x)^2]\log\frac{Q^2}{M^2}+(1-2x+2x^2)\log\left(\frac{1-x}{x}\right)-1+8x(1-x)\right]\label{ahg1dis}\\
A_{Hg}^{\DIS(2)}\left(x,\frac{Q^2}{M^2}\right)&=\frac{1}{x}\left[-8\log^2\frac{Q^2}{M^2}+\frac{80}{3}\log\frac{Q^2}{M^2}+\frac{40}{3}-\frac{8\pi^2}{3}\right]+{\cal O}(x^0).
\label{ahg2dis}
\end{align}
Taking the Mellin transforms $\mathbb{M}_N[A_{Hg}^{\DIS(1,2)}]$ then the limit $N\rightarrow0$ reproduces the terms in equation (\ref{ahg4}). Thus the LL terms in the heavy flavour matrix element are correctly predicted by the small $x$ resummation.\\
It is instructive to examine the behaviour of $A_{Hg}$ when $Q^2=M^2$, as this governs the initial value of the charm distribution at the matching point. It may be seen that the term in $\alpha_S^2$ in equation (\ref{ahg3}) is negative at $Q^2=M^2$, potentially driving the entire matrix element negative as $N\rightarrow 0$ resulting in a charm distribution which does not turn on with a positive value. This is not formally a problem given the interpretation of parton distributions as probability densities is only true in the na\"{i}ve parton model. Such a situation already occurs (and can be dealt with \cite{Chuvakin,Thorne06}) in NNLO perturbation theory in the $\msbar$ scheme, where the ${\cal O}(\alpha_S)^2$ contribution to $A_{Hg}$ \cite{Buza2} ensures that the charm turns on with a negative value. However, one would hope in a more physically motivated scheme such as the $\DIS(\chi)$ scheme that the partons are positive, due to their being related at zeroth order to the heavy flavour structure function $F_{2,H}$.\\

\begin{figure}
\begin{center}
\scalebox{0.9}{\includegraphics{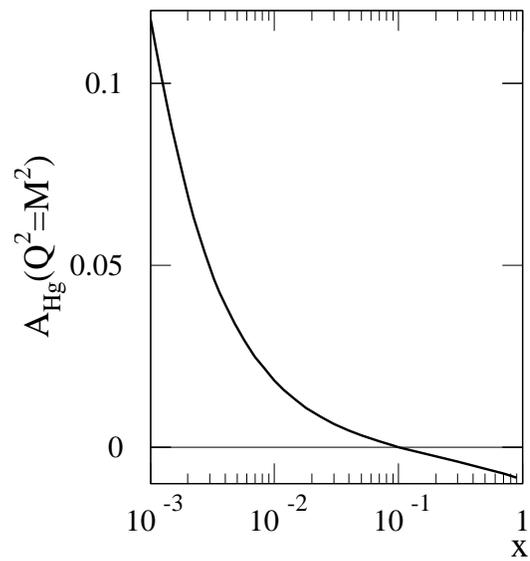}}
\caption{The heavy flavour matrix element $A_{Hg}$ at LL order, shown for $Q^2=M^2=2.25\text{GeV}^2$.}
\label{ahgplot}
\end{center}
\end{figure}

In fact, one may resolve this by considering higher orders in the fixed order expansion at small $x$. Evaluating the numerical coefficients of the leading logarithmic terms yields:
\begin{align}
A_{Hg}|_{Q^2=M^2}&=a\left[.66667-12.986\frac{a}{N}+267.96\left(\frac{a}{N}\right)^2-2794.8\left(\frac{a}{N}\right)^3+31076\left(\frac{a}{N}\right)^4\right.\notag\\
&\left.\qquad+1946155\left(\frac{a}{N}\right)^5+{\cal O}(a^6)\right],
\label{ahgseries}
\end{align}
where $a=\alpha_S/(4\pi)$. One sees that not every term in the fixed order expansion is negative. At ${\cal O}(\alpha_S^3)$, the complete matrix element becomes positive again in the fixed order expansion at small $x$. The LL order resummed matrix element expanded to ${\cal O}(\alpha_S^{17})$ (a very good approximation to the all orders result for $x\geq10^{-5}$) at $Q^2=M^2$ for $M=1.5$GeV is shown in figure \ref{ahgplot}. 
Although slightly negative at higher $x$, the small $x$ limit is positive and sets in at even intermediate values of $x$. One must also bear in mind the $\delta(1-x)$ contribution, absent in the plot, which is also positive. Indeed, the matrix element is positive for all values of $N$ in Mellin space. The positivity as $x\rightarrow 0$ is in contrast to the conclusion reached at ${\cal O}(\alpha_S^2)$ in the fixed order expansion. This highlights the importance of the high energy resummation, and it is reassuring that such a physical scheme choice results in a positive heavy quark contribution to the structure function.\\

\section{Resummed Heavy Flavour Coefficients}
\subsection{The BFKL Equation with Running Coupling}
So far we have presented the $\DIS(\chi)$ scheme at fixed order, and in double Mellin space at small $x$. Explicit results for the $x'$-space heavy flavour coefficients can only be obtained after solving the BFKL equation for the gluon density in equation (\ref{highfactor}). In this section we adopt the method of solution of \cite{Thorne01}, which is at LL order but includes the running of the coupling $\alpha_S$. Although formally a NLL effect, including the running coupling changes the nature of the solution of the BFKL equation and helps to stabilise the small $x$ expansion. \\

A comment is in order regarding the use of resummed coefficient functions and quark-gluon splitting functions in a LL fit. The impact factors calculated at LL order actually contribute to the coefficient and splitting functions at NLL level (n.b. terms of order $\alpha_S(\alpha_S/N)^n$ in Mellin space), according to the definition of e.g. \cite{Catani97}. Only the anomalous dimensions $\gamma_{gg}$ and $\gamma_{gq}$ contain LL terms, which arise from the LL BFKL kernel. We adopt the viewpoint that the leading resummations should be present in each physical quantity, and thus include the effects of the impact factors in a fit to data. After all, the quark-gluon splitting function and coefficient functions are zero in the strict definition of the LL expansion. Not including the NLL terms in these is analogous to disregarding the evolution of the partons in a LO fit in fixed order perturbation theory due to the fact that only the quark coefficient is non-zero at ${\cal O}(\alpha_S^0)$. This is nonsensical if LO QCD is to differ from the na\"{i}ve parton model. Likewise, if resummed perturbation theory is to differ from the LO fit, in both the parton coefficients and the evolution, one must include the impact factors. \\

We now recap the essential features of the method. A full presentation and discussion can be found in \cite{Thorne01}. When the LO running coupling is used, the BFKL equation in double Mellin space becomes a first order differential equation \cite{Collins2}:
\begin{equation}
\frac{df(\gamma,N)}{d\gamma}=\frac{df_I(\gamma,Q_0^2)}{d\gamma}-\frac{3}{\pi\beta_0N}\chi_0(\gamma)f(\gamma,N),
\label{BFKL3}
\end{equation}
where $f_I(\gamma,Q_0^2)$ is a non-perturbative input gluon 4-point function, usually taken to be to be a delta function in momentum space $\delta(k^2-Q_0^2)$. Also, $\chi_0$ is the Lipatov function:
\begin{equation}
\chi_0(\gamma)=2\psi(1)-\psi(\gamma)-\psi(1-\gamma).
\label{chi}
\end{equation}
Solving, one obtains:
\begin{equation}
f(\gamma,N)=\exp{[-X_0(\gamma)/(\bar{\beta}_0N)]}\int_\gamma^\infty\frac{df_I(\tilde{\gamma},Q_0^2)}{d\tilde{\gamma}}\exp{[X_0(\tilde{\gamma})/(\bar{\beta_0}N)]}d\tilde{\gamma},
\label{sol1}
\end{equation}
with $\bar{\beta}_0=\pi\beta_0/3$, and:
\begin{equation}
X_0(\gamma)=\int_{1/2}^\gamma\chi_0(\tilde{\gamma})d\tilde{\gamma}=\left[2\psi(1)\left(\gamma-\frac{1}{2}\right)-\ln{\left(\frac{\Gamma(\gamma)}{\Gamma(1-\gamma)}\right)}\right].
\label{X_0}
\end{equation}
Up to ${\cal O}(Q_0^2/Q^2)$ corrections, the integral in equation (\ref{sol1}) can be taken from 0 to $\infty$ \cite{Thorne01}, which leads to factorisation, up to power corrections, of the integrated gluon distribution as follows. Including the bare gluon as before, the integrated gluon structure function is obtained by inverse Mellin transformation of equation (\ref{sol1})\footnote{A factor of $\gamma^{-1}$ now arises from considering the integrated gluon structure function. The unintegrated function is related to this by a derivative with respect to $\ln{Q^2}$ corresponding to a factor of $\gamma$ in Mellin space.}:
\begin{equation}
{\cal G}(Q^2,N)=\int_{1/2-\imath\infty}^{1/2+\imath\infty}\frac{d\gamma}{2\pi\imath}\frac{1}{\gamma}\left(\frac{Q^2}{\Lambda^2}\right)^\gamma f(\gamma,N)g_B(Q_0^2,N).
\label{solinv}
\end{equation}
The Mellin transform of $f_I=\delta(k^2-Q_0^2)$ with respect to $k^2/\Lambda^2$ is $(Q_0^2/\Lambda^2)^{-\gamma-1}$. Inserting this into equation (\ref{solinv}) yields:
\begin{equation}
{\cal G}(Q^2,N)={\cal G}_E(Q^2,N){\cal G}_I(Q_0^2,N)g_B(Q_0^2,N),
\label{sol2}
\end{equation}
where :
\begin{equation}
{\cal G}_E(Q^2,N)=\frac{1}{2\pi\imath}\int^{1/2+\imath\infty}_{1/2-\imath\infty}\frac{1}{\gamma}f^{\beta_0}\exp{[\gamma\ln{(Q^2/\Lambda^2)}-X_0(\gamma)/(\bar{\beta}_0N)]}d\gamma
\label{G_E}
\end{equation}
and:
\begin{equation}
{\cal G}_I(Q_0^2,N)=\int_0^\infty\exp{[-\tilde{\gamma}\ln{(Q_0^2/\Lambda^2)}+X_0(\tilde{\gamma})/(\bar{\beta}_0 N)]}d\tilde{\gamma}.
\label{G_I}
\end{equation}
Note that a $\tilde{\gamma}$-independent factor of $-(\Lambda^2/Q_0^2)\ln{(Q_0^2/\Lambda^2)}$ has been absorbed into $g_B(Q_0^2,N)$. The function ${\cal G}_I$ is not calculable due to the presence of branch points on the contour of integration, but evolution of the gluon distribution depends only on the perturbative piece ${\cal G}_E$. The latter has been modified by a factor originating from the choice of renormalisation scale of $\alpha_S$ \footnote{See \cite{Thorne01} for a discussion.}:
\begin{equation}
f^{\beta_0}=\exp[1/2(\ln(\chi_0(\gamma))+X_0(\gamma))].
\label{fb0}
\end{equation}
To calculate a structure function, one follows the above derivation but includes an impact factor $h_a(\gamma)$, to yield the perturbative contribution (to be combined with the gluon distribution):
\begin{equation}
{\cal F}_{E,a}=\frac{1}{2\pi\imath}\int^{1/2+\imath\infty}_{1/2-\imath\infty}\frac{h_a(\gamma,Q^2/M^2)}{\gamma}f^{\beta_0}(\gamma)\exp{[\gamma\ln{(Q^2/\Lambda^2)}-X_0(\gamma)/(\bar{\beta}_0N)]}d\gamma.
\label{F_E}
\end{equation}
For the longitudinal coefficients and $F_{2,H}$ FF coefficient, one uses the factors $h_L$ and $h_2$ respectively. For the $F_{2,H}$ VF coefficient, one uses the result of equation (\ref{c2gvf}). Coefficient functions are now obtained via:
\begin{equation}
C_{a,g}^{FF,VF}(\alpha_S(Q^2),N)=\frac{{\cal F}_{E,a}(N,t)}{{\cal G}_E(N,t)}.
\label{coeffs}
\end{equation}
The integrals can be calculated numerically, but approximate analytic expressions may be obtained by expanding the Lipatov function:
\begin{equation}
\chi_0(\gamma)=\frac{1}{\gamma}+\sum_{n=1}^{\infty}2\zeta(2n+1)\gamma^{2n},
\label{Lipatov2}
\end{equation}
valid for $-1<\gamma<1$. One then finds, by deforming the integration contour of equation (\ref{F_E}) to enclose the negative real axis, and taking into account the discontinuity across the cuts \cite{Thorne01}:
\begin{equation}
\begin{split}
{\cal F}_{E,a}(N,t)=-\sin{\left(\frac{\pi}{\bar{\beta}_0N}\right)}\exp{\left(-\frac{\gamma_E}{\bar{\beta}_0N}\right)}\int_{-\infty}^0 f^{\beta_0}h_a(\gamma,Q^2/M^2)\gamma^{-1/(\bar{\beta}_0N)-1} \\  \exp{\left(\gamma t-\frac{1}{\bar{\beta}_0N}\sum_{n=1}^\infty a_n\gamma^{2n+1}\right)}d\gamma,
\end{split}
\label{F_E2}
\end{equation}
where $t=\ln{(Q^2/\Lambda^2)}$, $a_n=2\zeta(2n+1)/(2n+1)$, and $\gamma_E$ is Euler's constant. There is a correction of order $\Lambda^2/Q^2$ to equation (\ref{F_E2}), given that equation (\ref{Lipatov2}) is not valid for $\gamma\leq -1$. To evaluate this integral, one first changes variables to $y=\gamma t$, and then Taylor expands in $y$ the factor:
\begin{equation}
h_a\left(\frac{y}{t},\frac{Q^2}{M^2}\right)f^{\beta_0}\left(\frac{y}{t}\right)\exp{\left[\frac{1}{\bar{\beta}_0N}\sum_{n=1}^\infty a_n\left(\frac{y}{t}\right)^{2n+1}\right]}\equiv \sum_{n=0}K_{a,n}(\bar{\beta}_0,N) \left(\frac{y}{t}\right)^n.
\label{Taylor}
\end{equation}
Then one has:
\begin{equation}
\begin{split}{\cal F}_{E,a} = -\sin{\left(\frac{\pi}{\bar{\beta}_0N}\right)}\exp{\left(\frac{\gamma_E}{\bar{\beta}_0N}\right)}t^{1/(\bar{\beta}_0N)}\\ \times \sum_nK_{a,n}(\bar{\beta}_0,N)t^{-n}\int_{-\infty}^0y^{-1/(\bar{\beta}_0N)-1}\exp{(y)}y^n.
\end{split}
\label{Taylor2}
\end{equation}
Each of the integrals in the sum can be evaluated using:
\begin{equation}
\int_{-\infty}^0dy y^{-1/(\bar{\beta}_0)-1}\exp{(y)}y^n=(-1)^{-1/(\beta_0N)-1}\Delta_n(-1/(\bar{\beta}_0N))\Gamma(-1/(\bar{\beta}_0N)),
\label{integral}
\end{equation}
where:
\begin{equation}
\Delta_n(-1/(\bar{\beta}_0N))=\sum_{m=0}^{n-1}(-1)^md_{m,n}(\bar{\beta}_0N)^{-n+m},
\label{Delta}
\end{equation}
$d_{0,n}=1$, and $d_{m,n+1}=d_{m,n}+nd_{m-1,n}$. Setting $m=n$ in this recurrence relation yields $d_{n,n+1}=n d_{n-1,n}$, and hence:
\begin{equation}
d_{n-1,n}=(n-1)!,
\label{recur}
\end{equation}
and so the series in equation (\ref{Taylor2}) has factorially divergent coefficients $K_{a,n}(\bar{\beta}_0,N)$ at high $n$. Thus the series is asymptotic, and must be truncated at a finite order $n=n_0$. Finally, one has:
\begin{equation}
\begin{split}
{\cal F}_{E,a}=-\sin\left(\frac{\pi}{\bar{\beta}_0N}\right)\exp\left(-\frac{\gamma_E}{\bar{\beta}_0N}\right)\Gamma(-1/(\bar{\beta}_0N))t^{1/(\bar{\beta}_0N)}(-1)^{-1/(\beta_0N)-1}\\ \times\sum_{n=0}^{n_0}K_{a,n}(\bar{\beta}_0,N)(\bar{\alpha}_S\bar{\beta}_0)^n\Delta_n(-1/(\bar{\beta}_0N)),
\end{split}
\label{F_E3}
\end{equation}
using $t^{-n}=(\bar{\alpha}_S\bar{\beta}_0)^n$, $\bar{\alpha}_S=3\alpha_S/\pi$. A similar expression is found for the ${\cal G}_E$ (obtained from the above by setting the impact factor $h_a=1$), and then equation (\ref{coeffs}) is used to obtain the coefficient functions to any desired order in $\alpha_S$.\\

There is a slight caveat involving the resummed partons. In section 2, we defined the $\DIS(\chi)$ scheme at LL order, and gave the equivalent definition in the fixed order QCD perturbation expansion. However, this does not guarantee that the partons described in the two expansions are the same. Note that this is not a problem at leading order in $\alpha_S$, where all quantities are factorisation scheme independent. To be fully consistent with the conventional DIS scheme beyond LO, however, one must formulate the BFKL equation in $4+2\epsilon$ dimensions, regularising collinear singularities via dimensional regularisation. It does not presently seem possible to achieve phenomenological results once the running coupling is introduced, although a recent discussion may be found in \cite{Ciafaloni}. In the solution method used here, collinear factorisation is introduced in equation (\ref{sol1}) by modifying the lower limit of integration, with regularisation achieved by taking the gluon at the lower end of the BFKL ladder off-shell with virtuality $Q_0^2$. This certainly does not correspond to the way in which the fixed order expansion is factorised. We believe, however, that the formal difference between the partons in the fixed order and small-$x$ expansions can be neglected to a good approximation, and argue as follows. In the case of the LL BFKL equation with fixed coupling, solution by dimensional regularisation relates the gluon defined by solution of the BFKL equation (the so-called ``$Q_0$ scheme'') with the conventional $\msbar$ scheme gluon \cite{CCH}:
\begin{equation}
g^{Q_0}=R_Ng^{\msbar},
\label{Q_0}
\end{equation}
in $N$-space, where $R_N=1+{\cal O}(\alpha_S^3)$. Thus the difference between the partons in the two expansions is irrelevant when working up to ${\cal O}(\alpha_S^2)$ at fixed order. One hopes that running coupling effects do not seriously modify this situation. \\

A recent discussion of scheme transformation between the $Q_0$ and $\msbar$ schemes can be found in \cite{Ciafaloni05}, in which factorisation is derived in the dimensionally regularised BFKL equation (including the running coupling) using a saddle point approximation. The effect of the scheme transformation on a toy gluon distribution is investigated in \cite{Ciafaloni06}, where it is noted that the size of the correction in transforming from the $\msbar$ to $Q_0$ scheme is less significant than the difference between gluons obtained in the $\msbar$ scheme by two rival groups (MRST and CTEQ). Bearing in mind that the transformation discussed in \cite{Ciafaloni06} is more involved than the shift from the $Q_0$ scheme to a DIS-like scheme considered here, one is justified in assuming the change in parton normalisation to be a small effect. One must also consider the quark sector, which is more complicated given that the effect of a transformation from the $Q_0$ scheme enters at the same order of the high energy expansion as the leading resummation in the splitting function $P_{qg}$. We assume, however, that as in the gluon case the effect of transforming from the $Q_0$ scheme $\DIS(\chi)$ quark to the quark distribution obtained in the $\msbar$-subtracted $\DIS(\chi)$ scheme is small. \\

There are other minor technical problems with the procedure specified here. Firstly, the FF resummed coefficient function is evaluated with $\bb(n_f)$, whereas the resummed VF coefficient and heavy matrix elements $A_{Hg}$, $A_{Hq}$ are evaluated with $\bb(n_f+1)$. This results in a slight loss of continuity of $F_{2,H}$ across the matching point $Q^2=M^2$. We checked, however, that this is an extremely small effect. Secondly, the coupling is implemented in the solution of the BFKL equation using the simple LO form $\alpha_S=1/(\beta_0\log(k^2/\Lambda^2))$, whereas the implementation of the coupling when heavy flavours are accounted for is somewhat more complicated in that $\Lambda$ changes discontinuously as one crosses each heavy quark threshold. This we also expect to be a small effect, and less significant than other effects due to choice of scale \footnote{See the discussion regarding the term $f^{\beta_0}$ of our equation (\ref{fb0}) in \cite{Thorne01}.}.\\
\subsection{The Resummed Heavy Flavour Coefficients}
To follow this procedure for the heavy flavour coefficients, one must expand the impact factors as a series in $\gamma$. However, they contain hypergeometric functions in $(1+4M^2/Q^2)^{-1}$ \cite{CCH} and so one must first Taylor expand them in a suitable $Q^2$-dependent parameter. For low $Q^2$ one can use $z=(1+4M^2/Q^2)^{-1}$ and for high $Q^2$, $u=1-z$. One needs sufficiently many orders of $z$ or $u$ as are needed to give an accurate description of the impact factor over the range of interest. The impact factors relevant to the FF coefficients are expanded in terms of $z$, and two expressions are needed for the VF factors for $Q^2\geq M^2$ in terms of $z$ and $u$ respectively, overlapping at a suitable scale. We find that in the case of $C_{2,g}^{FF}$, 18 orders of $z$ are sufficient for a precision of a few parts permille up to $Q^2=M^2$. For $C_{2,g}^{VF}$, 9 orders in $z$ are needed up to $z=0.5$ ($Q^2=4M^2$), and then 9 orders of $u$ are sufficient for $z>0.5$. For $C_{L,g}^{FF,VF}$, 8 orders of both $z$ and $u$ are necessary, again matching at $z=0.5$. After these expansions, the resulting impact factors can be further expanded in $\gamma$. The series in $\gamma$ for ${\cal F}_E$ must ultimately be truncated due to its being an asymptotic series. How many orders to include can be decided by examining the ratio $\hat{h}_a/h_a$ over the range $\gamma\in(-1,0)$, where $\hat{h}_a$ is the double Taylor series in $z$ (or $u$) and $\gamma$. Contributions from $\gamma\leq 1$ are suppressed by inverse powers of the hard scale $Q^2$. This ratio is required to be at least as accurate as the similar ratio of $h$ factors for the massless case in \cite{Thorne01}. I find that 5 orders of $\gamma$ are needed for each of the impact factors. Although the resulting structure functions are of finite order in $\alpha_S$, the coefficients can be expanded to arbitrarily many orders in $\alpha_S/N$. Following \cite{Thorne01}, we require the series to be stable with respect to higher orders in $\alpha_S$ for $N\geq 0.4$ (corresponding to $x'\gtrsim 10^{-5}$). No significant improvement in any of the coefficients (to within 0.1\%) is found by taking more than 9 orders in $\alpha_S$.\\ 

The explicit expressions for the resummed $N$-space functions are extremely lengthy (owing in part to the high number of terms in $z$ or $u$ needed when expanding the impact factors), and so are not reproduced here. They have the general form:
\begin{equation}
C_{a,g}^{FF,VF}(\alpha_S,N,w)=\frac{\alpha_Sn_f}{4\pi}\sum_{n=0}^9\sum_{m=0}^{n+1}\bar{\alpha}_S^nC_{nm}(z)\frac{\bar{\beta}_0^{n-m}}{N^m},
\label{c2gform}
\end{equation}
with:
\begin{equation}
C_{nm}(z)=\sum_r\sum_s^5 B_{rs} w^r(\log {w})^s,
\label{cnm}
\end{equation}
and $w=z$ or $u$ as appropriate. Note the presence of terms $\sim\alpha_S^n/N^{n+1}$. These are strictly absent in the full coefficients, but arise here from the truncation of the asymptotic series for the structure functions in double Mellin space and are weighted by inverse powers of $\beta_0$. This is justified in that the aim is to produce accurate phenomenological representations of the resummed coefficient functions with running coupling effects, and thus one must truncate the series as specified. The leading logarithmic terms are the same as those that would arise if running coupling corrections were not included, up to the order of truncation. The $N$-space expressions are easily inverted to obtain the corresponding $x'$-space coefficient functions. For phenomenological applications, the coefficients $C_{nm}$ in equation (\ref{c2gform}) can be parameterised as a function of $y=\log{Q^2/4M^2}$. We used:
\begin{equation}
C_{nm}=\sum_{i=0}^5k_{nm}^iy^i.
\label{paramy}
\end{equation}
Several ranges of $y$ are needed to cover a $Q^2$ range from the parton starting scale 1 GeV up to a TeV with an accuracy of less than a percent, but the resulting parameterised expressions are much quicker than the exact results when used in a global fit.\\

\begin{figure}
\begin{center}
\scalebox{0.5}{\includegraphics{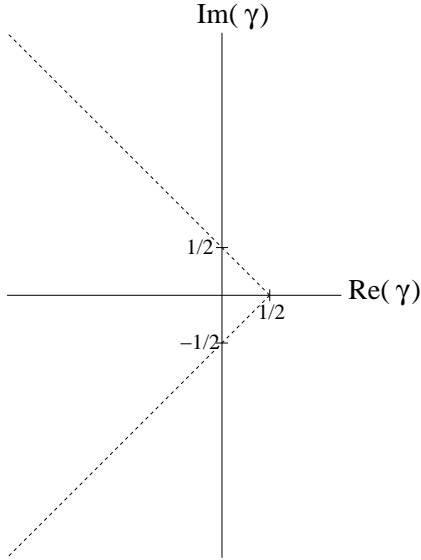}}
\caption{Contour for numerical integration of equations (\ref{G_E}) and (\ref{F_E}).}
\label{contour}
\end{center}
\end{figure}

No mention has yet been made of the power-suppressed correction to equation (\ref{Taylor2}) due to the ambiguities in its derivation. One can estimate this for each coefficient by comparing the analytically computed expression with the ratio of ${\cal F}_{E,a}$ and ${\cal G}_E$ as found by numerical integration of equations (\ref{G_E}) and (\ref{F_E}) in the $\gamma$-plane. Following \cite{Thorne01}, the contour for these numerical integrations is chosen as in figure \ref{contour}. This is a deformation of the contour used in the definition of the inverse Mellin transform (parallel to the imaginary axis in the $\gamma$ plane), and aids convergence of the numerical integration. One is allowed to make such a deformation given that branch points of the integrands of equations (\ref{G_E}, \ref{F_E}) occur only along the negative real axis, which can be used to set the direction of all cuts. No singularities are then encountered in bending the inversion contour to that of figure \ref{contour}.\\

\begin{figure}
\begin{center}
\scalebox{0.8}{\includegraphics{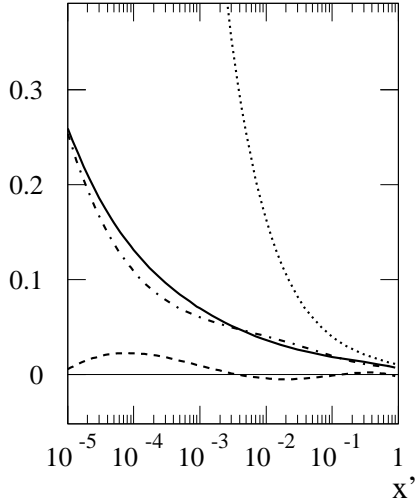}}
\caption{The total resummed coefficient $x'\,C_{2,g}^{FF}(\alpha_S,x',Q^2/M^2)$ for $t=4.61$ ($Q^2=M^2\simeq 2.25\text{GeV}^2$) and $n_f=3$ (solid line), together with the power-suppressed correction (dashed line) and perturbative piece (dot-dashed). Also shown is the LL correction with no running coupling corrections (dotted).}
\label{c2low}
\end{center}
\end{figure}

One can only perform the numerical integrations by choosing particular values of $M$ and $\Lambda$. I use $M=1.5$ GeV and $\Lambda=150$ MeV, where the heavy quark mass corresponds approximately to the charm quark. For each coefficient, the difference $\Delta C$ between the numerical and analytical results is found for $N\in\{0.4, 0.5, 0.6, 0.7, 0.8, 0.9, 1, 2, 5, \infty\}$ for two values $t=t_0$ and $t=t_1$, and then fitted to a function of the form:
\begin{equation}
\Delta C(N,t)=f(t)\exp{(-A t)}+\exp{(-Bt)}\sum_n b_n\left(\frac{\alpha_S(t)}{\alpha_S(t_0)}\right)^{c_n}\frac{1}{(N+d_n)^n}.
\label{error}
\end{equation}
The $N$-independent term is parameterised by finding $\Delta C$ at $N=\infty$ for various $t$ values in the range $t_0\leq t \leq t_1$. Usually a polynomial in $t$ is sufficient for $f(t)$. For $C_{2,g}^{FF}$ we use $t_0=3.8$ and $t_1=4.61$, corresponding at this $\Lambda$ to $Q^2=1\text{GeV}^2$ and $Q^2=M^2$ respectively. The $N$-independent term was found using $t\in\{3.8,3.85,3.9,4.0,4.1,4.2,4.3,4.4,4.5,4.61\}$, and the resulting expression for the $x'$-space correction is:
\begin{align}
x'\Delta C_{2,g}^{FF}&=\frac{\alpha_S}{4\pi}\left\{(-1.5874-.94368\log(t-3.6)+.33073t+.58757\log(t-3.7))e^{-t}\delta(1-x')\notag\right.\\
&\left.+e^{-t}\left[-11.523\left(\frac{\alpha_S(t)}{\alpha_S(3.8)}\right)^{-.2222558}+45.661\left(\frac{\alpha_S(t)}{\alpha_S(3.8)}\right)^{-.0311852}{\xi'}-66.891\left(\frac{\alpha_S(t)}{\alpha_S(3.8)}\right)^{.283791}\frac{{\xi'}^2}{2!}\right.\right.\notag\\
&\left.\left.+47.493\left(\frac{\alpha_S(t)}{\alpha_S(3.8)}\right)^{.749987}\frac{{\xi'}^3}{3!}-16.709\left(\frac{\alpha_S(t)}{\alpha_S(3.8)}\right)^{1.353423}\frac{{\xi'}^4}{4!}+2.3538\left(\frac{\alpha_S(t)}{\alpha_S(3.8)}\right)^{2.047850}\frac{{\xi'}^5}{5!}\right]\right\},
\label{powc2gff}
\end{align}
with ${\xi'}=\log(1/x')$. Note that non-polynomial terms in the $N$-independent contribution were needed for a reasonable fit with few parameters. A plot of the coefficient is shown in figure \ref{c2low} for $t=4.61$, together with the power-suppressed correction and the result at LL order with no running coupling corrections.\\

\begin{figure}
\begin{center}
\scalebox{0.8}{\includegraphics{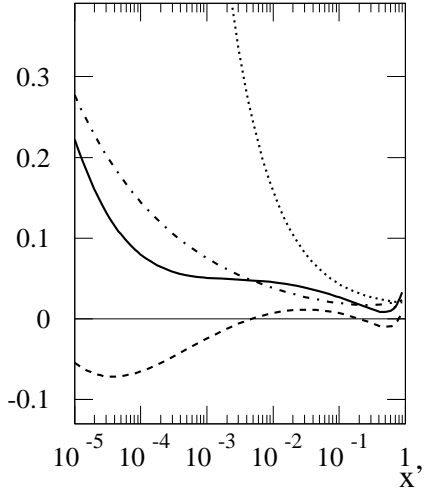}}
\caption{The total resummed coefficient $x'\,C_{2,g}^{VF}(\alpha_S,x',Q^2/M^2)$ for $t=4.61$ ($Q^2=M^2\simeq 2.25\text{GeV}^2$) and $n_f=4$ (solid line), together with the power-suppressed correction (dashed line) and perturbative piece (dot-dashed). Also shown is the LL result with no running coupling corrections (dotted).}
\label{c2high}
\end{center}
\end{figure}

For $C_{2,g}^{VF}$, we used $t_0=4.61$ and $t_1=7$, and the $N$-independent term was found by calculating the correction at $t \in \{4.61, 4.8, 5, 5.2, 5.4, 5.6, 5.8, 6, 6.2, 6.4, 6.6, 6.8, 7\}$. The power-suppressed correction in $x'$-space can be modelled by:
\begin{align}
&x'\Delta\,C_{2,g}^{VF} = \frac{\alpha_S}{4\pi}\left\{(-5.1401+4.3451t-.86521t^2+0.048160t^3)e^{-t}+e^{-t}\left[-172.07\left(\frac{\alpha_S(t)}{\alpha_S(4.61)}\right)^{2.862916}x'\right.\right.\notag\\
&+262.40\left(\frac{\alpha_S(t)}{\alpha_S(4.61)}\right)^{2.672627}x'^2+22.187\left(\frac{\alpha_S(t)}{\alpha_S(4.61)}\right)^{3.25314}{\xi'}-7.1036\left(\frac{\alpha_S(t)}{\alpha_S(4.61)}\right)^{3.811813}\frac{{\xi'}^4}{4!}\notag\\
&\left.\left.+3.9345\left(\frac{\alpha_S(t)}{\alpha_S(4.61)}\right)^{4.02541}\frac{{\xi'}^5}{5!}-.58657\left(\frac{\alpha_S(t)}{\alpha_S(4.61)}\right)^{4.24327}\frac{{\xi'}^6}{6!}\right]\right\},
\label{powc2gvf}
\end{align}
and is shown with the variable flavour coefficient in figure \ref{c2high}.\\

\begin{figure}
\begin{center}
\scalebox{0.8}{\includegraphics{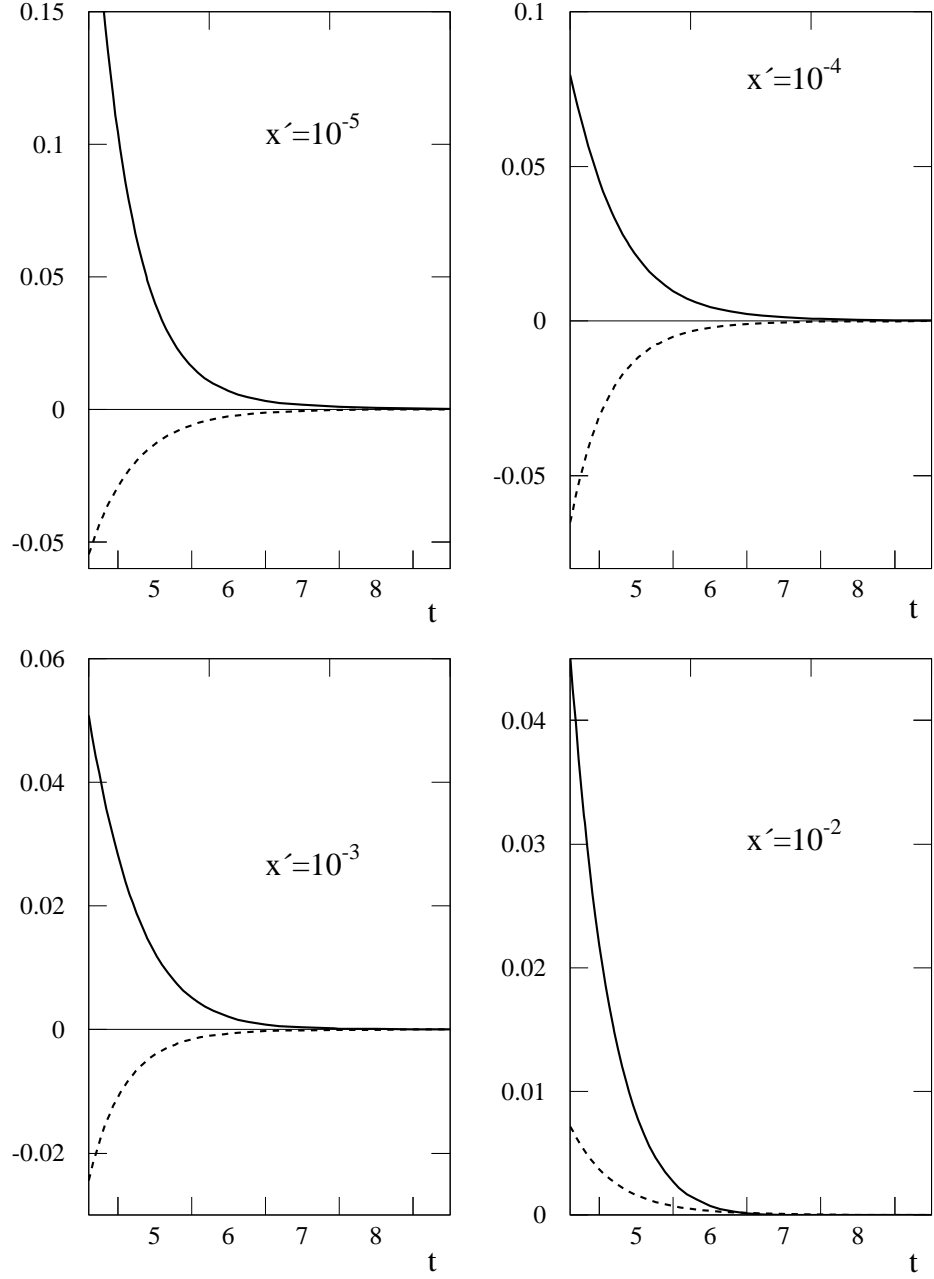}}
\caption{The total resummed coefficient $x'C_{2,g}^{VF}$ as a function of $t=\log(Q^2/\Lambda^2)$ at various values of $x'$ for $n_f=4$ (solid lines), shown with the power-suppressed correction (dashed lines).}
\label{c2t}
\end{center}
\end{figure}

A plot of the VF coefficient as $t$ increases is shown in figure \ref{c2t}.
Note that the power-suppressed correction remains a non-negligible correction to $C_{2,g}^{VF}$ even for intermediate $x'$ values and higher $t$. This is due to the fact that both the correction and the coefficient are decaying as $t\rightarrow\infty$, where both quantities are becoming less phenomenologically important.\\

\begin{figure}
\begin{center}
\scalebox{0.8}{\includegraphics{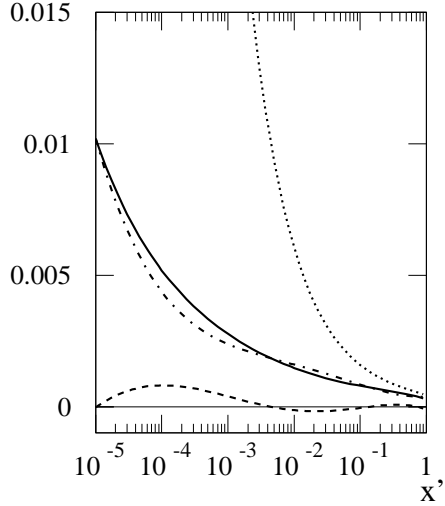}}
\caption{The $x'$-space coefficient $x'\,C_{L,g}^{VF}$ for $t=4.5$ and $n_f=4$ (solid line), together with the power-suppressed correction (dashed) and perturbative piece (dot-dashed). Also shown is the LL result with no running coupling corrections (dotted).}
\label{clgplot}
\end{center}
\end{figure}

For the longitudinal coefficients, we used $t_0=4.5$ and $t_1=6$ (the correction dies away relative to the coefficient more rapidly in the longitudinal case, as $C_{L,g}^{VF}$ does not vanish as $Q^2/M^2\rightarrow\infty$). The $N$-independent term was found by considering $t\in\{4.5, 4.75, 5, 5.25, 5.5, 5.75, 6\}$. The result is:
\begin{align}
x' \Delta C_{L,g} &= \frac{\alpha_S}{4\pi}\left\{(-5.1401+4.3451t-.86521t^2+0.048160t^3)e^{-t}+e^{-t}\left[-.34041\left(\frac{\alpha_S(t)}{\alpha_S(4.5)}\right)^{-6.00488}\right.\right.\notag\\
&+1.3616\left(\frac{\alpha_S(t)}{\alpha_S(4.5)}\right)^{-5.27119}{\xi'}-1.9545\left(\frac{\alpha_S(t)}{\alpha_S(4.5)}\right)^{-4.76871}\frac{{\xi'}^2}{2!}+1.3184\left(\frac{\alpha_S(t)}{\alpha_S(4.5)}\right)^{-4.27245}\frac{{\xi'}^3}{3!}\notag\\
&\left.\left.-.42836\left(\frac{\alpha_S(t)}{\alpha_S(4.61)}\right)^{-3.72971}\frac{{\xi'}^4}{4!}+0.054590\left(\frac{\alpha_S(t)}{\alpha_S(4.5)}\right)^{-3.12486}\frac{{\xi'}^5}{5!}\right]\right\}
\label{powclg}
\end{align}
(n.b. this is the correction to the FF coefficient for $Q^2\leq M^2$, and the VF coefficient for $Q^2\geq M^2$). The coefficient and power-suppressed correction are shown in figure \ref{clgplot} for $t=4.5$. We note that the shape is very similar to $C_{2,g}^{FF}$ (figure \ref{c2high}). There are also resummations contributing to the heavy quark coefficient $C_{L,HH}=C_{L,q}^{\DIS}(x')$, where the form of the massless resummed quark coefficient can be found in \cite{Thorne01}.\\

Once the gluon coefficients are known, the resummed coefficients to be convolved with the singlet distribution are calculated via the colour relation:
\begin{equation}
C_{2,Hq}=\frac{C_F}{C_A}[C_{2,Hg}-C_{2,Hg}^{(1),rs}],
\label{colourcharge}
\end{equation}
where $C_F=4/3$, $C_A=N_c=3$, and $C_{2,Hg}^{(1),rs}$ denotes the ${\cal O}(\alpha_S)$ contribution from the resummed gluon coefficient. This arises from the coupling of quarks to the BFKL 4-point function at the lower end, and is true to LL order which is consistent with the order of the resummation.\\

From figures \ref{c2low}-\ref{clgplot}, one sees that the effect of the running coupling is generally to suppress the high energy limit until lower values of $x$, and also to moderate the small $x$ divergence. It is interesting to compare our results with those of alternative analytic approaches. In \cite{Salam1}, the BFKL kernel is improved by requiring consistency of the small $x$ expansion with the DGLAP limit of the gluon Green's function. A resummation of scale-dependent collinear singularities of the kernel \cite{Salamcp} is implemented, and a generalised BFKL equation presented in which the kernels at each order of $\alpha_S$ depend on both $N$ and $\gamma$ in double Mellin space. A method for solving this generalised equation (the ``$\omega$-expansion'' in the notation of \cite{Salam1}) is given. This is further developed in \cite{Salam2}, where results are presented for the gluon Green's function and gluon-gluon splitting function, as well as a discussion of impact factors needed for detailed phenomenology. A different approach is taken in \cite{ABF1}, where the BFKL anomalous dimension and kernel are improved using a duality relation between the two quantities derived on the assumption that the rightmost singularities of the gluon in the $\gamma$- and $N$- planes coincide. The duality relation can be generalised to the running coupling case, and a proof that this can be extended to all orders was given recently \cite{BF}. Further improvements arise from the imposition of momentum conservation in the gluon evolution, and the resummation of collinear poles \cite{Salamcp} - see \cite{ABF} for a recent discussion. As far as phenomenology is concerned, however, there is no discussion of the incorporation of impact factors in the duality approach, which are needed for $P_{qg}$ and the resummed coefficient functions. \\

These approaches to the BFKL expansion are different to the method used here and in principle more complicated, but it is interesting to note that all three share the same qualitative results. The inclusion of the running coupling, present in all the approaches but most transparent in this one, stabilises the resummed perturbation theory and significantly moderates the small $x$ divergence. Also common to each approach is the presence of a dip in the splitting function $P_{gg}$ at intermediate $x$ (see figure 6 of \cite{Thorne01} for the result adopting the method used here), before the small $x$ limit sets in. Indeed, in all physical quantities the asymptotic small $x$ behaviour seems to be pushed to lower values of $x$.
\subsection{The Heavy Flavour Matrix Element}
\begin{figure}
\begin{center}
\scalebox{0.8}{\includegraphics{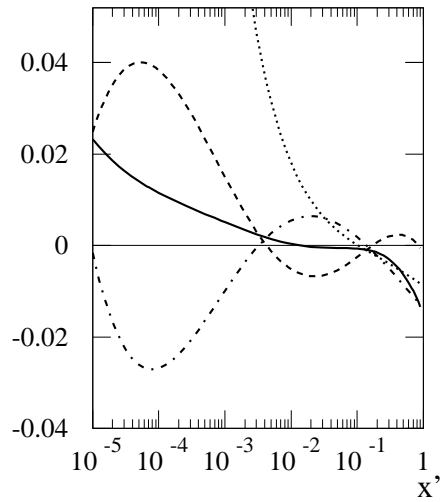}}
\caption{The total resummed matrix element $A_{Hg}$ with running coupling corrections for $n_f=4$ (solid line), together with the power-suppressed correction (dashed) and perturbative piece (dot-dashed). Also shown is the LL result with no $\beta_0$-dependent corrections (dotted).}
\label{ahgx}
\end{center}
\end{figure}

In order to implement the small $x$ resummations in a fit to data, one also needs the heavy flavour matrix element $A_{Hg}$ with running coupling corrections, evaluated at the matching scale. One thus follows the procedure above, using an ``h-factor'' given by equation (\ref{ahg2}) with $Q^2=M^2$. We again find the series in double Mellin space can be reliably truncated at 5 orders in $\gamma$, and the result for the resummed matrix element is:
\begin{align}
xA_{Hg}&=\frac{\alpha_S}{4\pi}\left[.66667\delta(1-x)-1.0821\bar{\alpha}_S+\left(1.8609\xi-1.8609\bb\right)\abar^2+\left(-1.6174\frac{\xi^2}{2!}+4.8521\bb\xi\right.\right.\notag\\
&\left.-3.2347\bb^2\right)\abar^3+\left(1.4987\frac{\xi^3}{3!}-11.594\bb\frac{\xi^2}{2!}+24.290\bb^2\xi-14.195\bb^3\right)\abar^4+\left(7.8212\frac{\xi^4}{4!}\right.\notag\\
&\left.-51.369\bb\frac{\xi^3}{3!}+121.63\bb^2\frac{\xi^2}{2!}-122.63\bb^3\xi+44.549\bb^4\right)\abar^5+\left(-1.7449\frac{\xi^6}{\bb6!}+16.204\frac{\xi^5}{5!}\right.\notag\\
&\left.-61.577\bb\frac{\xi^4}{4!}+118.82\bb^2\frac{\xi^3}{3!}-114.39\bb^3\frac{\xi^2}{2!}+42.685\bb^4\xi\right)\abar^6+\left(1.9728\frac{\xi^7}{\bb7!}-26.101\frac{\xi^6}{6!}\right.\notag\\
&\left.+135.73\bb\frac{\xi^5}{5!}-357.64\bb^2\frac{\xi^4}{4!}+503.70\bb^3\frac{\xi^3}{3!}-358.74\bb^4\frac{\xi^2}{2!}+101.07\bb^5\xi\right)\abar^7+\left(5.5968\frac{\xi^8}{\bb8!}\right.\notag\\
&\left.-61.771\frac{\xi^7}{7!}+282.34\bb\frac{\xi^6}{6!}-687.81\bb^2\frac{\xi^5}{5!}+955.67\bb^3\frac{\xi^4}{4!}-748.17\bb^4\frac{\xi^3}{3!}+300.36\bb^5\frac{\xi^2}{2!}\right.\notag\\
&\left.-46.201\bb^6\xi\right)\abar^8+\left(-1.3983\frac{\xi^{10}}{\bb^210!}+20.599\frac{\xi^9}{\bb9!}-138.35\frac{\xi^8}{8!}+564.42\bb\frac{\xi^7}{7!}-1560.3\bb^2\frac{\xi^6}{6!}\right.\notag\\
&\left.\left.+3036.6\bb^3\frac{\xi^5}{5!}-4111.0\bb^4\frac{\xi^4}{4!}+3652.2\bb^5\frac{\xi^3}{3!}-1886.7\bb^6\frac{\xi^2}{2!}+423.92\bb^7\xi\right)\abar^{10}\right],
\label{ahgrs}
\end{align}
where $\xi=\log(1/x)$. The power-suppressed correction is somewhat easier for this quantity, as it is only needed for one value of $t$. The difference $\Delta A$ between the analytical and numerically evaluated results is found for $N\in\{0.4, 0.5, 0.6, 0.7, 0.8, 0.9, 1, 2, 5, \infty\}$ and fitted to a function of $N$ only. The result in $x$-space is then:
\begin{equation}
x\Delta A=-0.013550\delta(1-x)-0.058138+.39486\xi-.62386\frac{\xi^2}{2!}+.41538\frac{\xi^3}{3!}-.12522\frac{\xi^4}{4!}+0.014138\frac{\xi^5}{5!}.
\label{ahgpow}
\end{equation}
A plot of the matrix element is shown in figure \ref{ahgx}, alongside the corresponding LL order result evaluated to ${\cal O}(\alpha_S^9)$. 
Note that the power-suppressed correction dominates the analytic result. This problem is made worse if one includes more powers of $\gamma$ in double Mellin space due to the instability of the asymptotic expansion. Neither is the size of the correction reduced by expanding the analytic result to higher powers in $\alpha_S$. However, this is not really a problem given that the power-suppressed correction is modelled very well due to only being required at a single value of $Q^2$. One can see from figure \ref{ahgx} that the effect of the correction is to restore positivity of the matrix element at small $x$, as noted in $N$-space previously. We again note that the running coupling corrections moderate the small $x$ divergence from the strictly LL result. The resummed singlet matrix element $A_{Hq}$ is calculated from $A_{Hg}$ analogously to equation (\ref{colourcharge}).

\section{Global fit to Scattering Data}
\subsection{Data Sets and Details of the Fit}
In this section we present the results of a global parton fit adopting the $\DIS(\chi)$ factorisation scheme outlined in the preceding sections, implementing the resummed splitting functions in the parton evolution and the resummed coefficient functions in the theoretical $F_2$ and $F_L$ predictions for electromagnetic current DIS. We include structure function data from the H1 \cite{H1a,H1b,H1c} and ZEUS \cite{ZEUSa,ZEUSb,ZEUSc} collaborations at HERA; proton data from BCDMS \cite{BCDMSep}, NMC \cite{NMC}, SLAC \cite{SLAC,SLAC2} and E665 \cite{E665}; deuterium data from BCDMS \cite{BCDMSeD}, NMC, SLAC and E665; CCFR data on $F_{2,3}^{\nu(\bar{\nu})N}(x,Q^2)$ \cite{CCFRF2, CCFRF3}; data on the deuterium-proton ratio $F_2^D/F_2^p$ from NMC \cite{NMCrat}; charged current data from H1 \cite{H1a} and ZEUS \cite{ZEUSCC}; data on the charm structure function $F_{2,c}$ from H1 \cite{F2c1} and ZEUS \cite{F2c2}. The non-DIS data sets used are Drell-Yan (DY) data from the E866/NuSea collaboration \cite{DY}; DY asymmetry data from NA51 \cite{DYasym}; data on the DY ratio $\sigma_{DY}^{pD}/\sigma_{DY}^{pp}$ from E866 \cite{DYrat}; W-asymmetry data from CDF \cite{Wasym}. By means of a consistency check, we also compare the resulting gluons with a gluon from a previous NLO fit, which includes Tevatron jet data from the D0 \cite{Tevjet1} and CDF \cite{Tevjet2} collaborations. \\

The parton distributions are parameterised using the forms given in \cite{MRST2001} at a starting scale of $Q_{init}^2=1\text{GeV}^2$, allowing for a valence-like or even negative gluon at the starting scale. The running coupling is implemented according to the prescription of \cite{Marciano} for dealing with heavy flavour thresholds. As well as including small $x$ resummations as outlined in this paper, we also include a resummation of the leading large $x$ divergences in the anomalous dimension $P_{qq}$ by modifying the argument of $\alpha_S$ acting on the plus distribution from $Q^2\rightarrow Q^2(1-x)$ \cite{Amati}. As will be seen, such a resummation is needed in order to achieve agreement between theory and data at high $x$. To avoid encountering the Landau pole in the coupling as a result of this modification, $\alpha_S$ is frozen for $Q^2\leq0.144\text{GeV}^2$ ($\alpha_S\sim 0.6$).\\

A problem occurs with the Drell-Yan data in that the NLO QCD corrections to the cross-section are known to be large \cite{AltarelliDY}. To obtain a reasonable fit to the Drell-Yan data, we thus employ a constant ``K-factor'' ($\sim 1.4$) as in \cite{MRSTLO}, whose effect is normalise the DY data outside the confines of the systematic error. Although strictly absent at LO, such a factor does not have any $x_1$, $x_2$ or $Q^2$ dependence and thus only poorly models the known higher order corrections.\\

\subsection{Results}
In table \ref{LLfit2}, we show the $\chi^2$ values for each data set obtained in the LL fit, together with the results obtained from a LO fit with no resummations at small or large $x$.\\
\begin{table}
\begin{center}
\begin{tabular}{|cc|ccc|}
\hline
Data Set & No. data pts & $\chi^2_{LL}$ & $\chi^2_{LO}$ & $\chi^2_{NLO}$ \\
\hline
H1 ep & 417 &342 &414 & 427\\
ZEUS ep &356 &282 &287 &279\\
$F_2^c$ &27 &26 &24&32\\
BCDMS $\mu$ p &167 &170 &263&191\\
BCDMS $\mu$ D &155 &230 &208&216\\
NMC $\mu$ p &126 &111 &154&136\\
NMC $\mu$ D &126 &89 &141&103\\
SLAC ep &53 &77 &195&67\\
SLAC eD &54 &74 &188&56\\
E665 $\mu$ p &59 &59 &56&61\\
E665 $\mu$ D &57 &55 &54&51\\
CCFR $F_2^{\nu N}$ &74 &99 &158&83\\
CCFR $F_3^{\mu N}$ &105 &138 &126&115\\
H1 CC & 28 &33 &34 &29\\
ZEUS CC &30 &46 &34 &35\\
NMC $n/p$ &156 &164 &144&154\\
E866/ NuSea DY &174 &307 &276&237\\
NA51 DY asym. &1 &11 &6&11\\
E866 $\sigma_{DY}^{pD}/\sigma_{DY}^{pp}$ &15 &7 &29&10\\
CDF $W$ asym. &11 &16 &26&14\\
\hline
Total &2181 &2336 &2817& 2307\\
\hline 
\end{tabular}
\caption{The quality of fit from the LL and LO fits for each dataset, as well as results from a previous NLO fit.}
\label{LLfit2}
\end{center}
\end{table}

The LO fit is reasonable, but misleading. It fails due to the fact that the evolution is too slow. Some compensation is achieved by increasing the QCD scale parameter $\Lambda$, in that by raising the coupling constant $\alpha_S$ the quarks evolve faster, and one can increase the theoretical predictions to be more in line with the data. The LO fit gives $\alpha_S(M_Z^2)=0.1305$, which is indeed rather high when compared to the world average of $0.1187(20)$ \cite{PDG}. This is not strictly a problem at LO, where one is free to redefine the renormalisation scale - chosen here to be $\mu_R^2=Q^2$. It is still, however, a cause for concern given that $Q^2$ is the natural scale choice. As well as a raised coupling constant, the H1 and ZEUS normalisations are required to be lower in the LO fit than the corresponding resummed fit parameters. The description of $F_2$ at small and high $x$ fails ultimately because the shape is too flat. The fit to the SLAC data, for example, improves significantly once high $x$ resummations are included. \\

\begin{figure}
\begin{center}
\scalebox{0.7}{\includegraphics{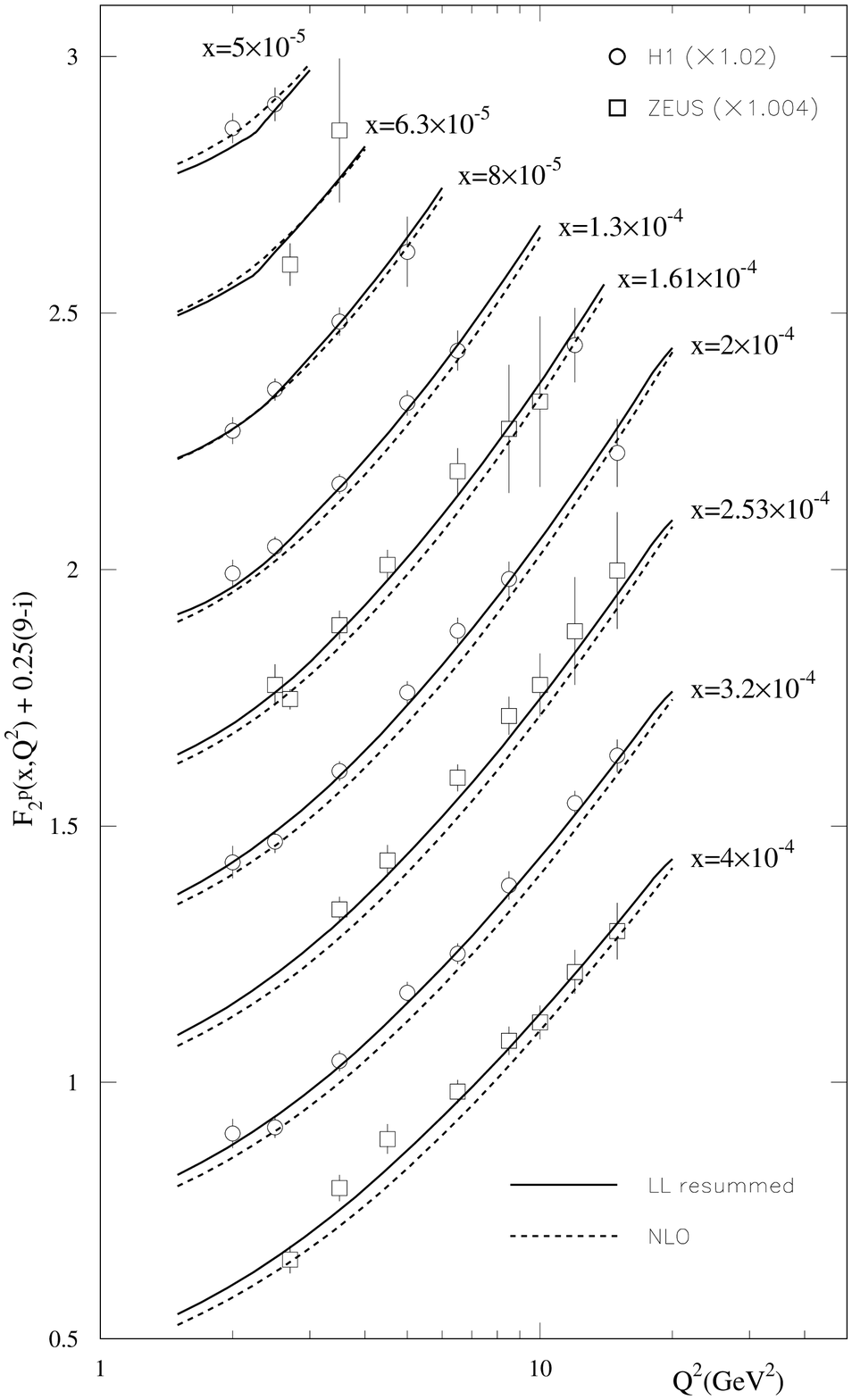}}
\caption{Theoretical predictions for the structure function $F_2$ alongside the data (normalisation dictated by the fit), for $5\times10^{-5}\leq x\leq4\times10^{-4}$.}
\label{data}
\end{center}
\end{figure}

\begin{figure}
\begin{center}
\scalebox{0.7}{\includegraphics{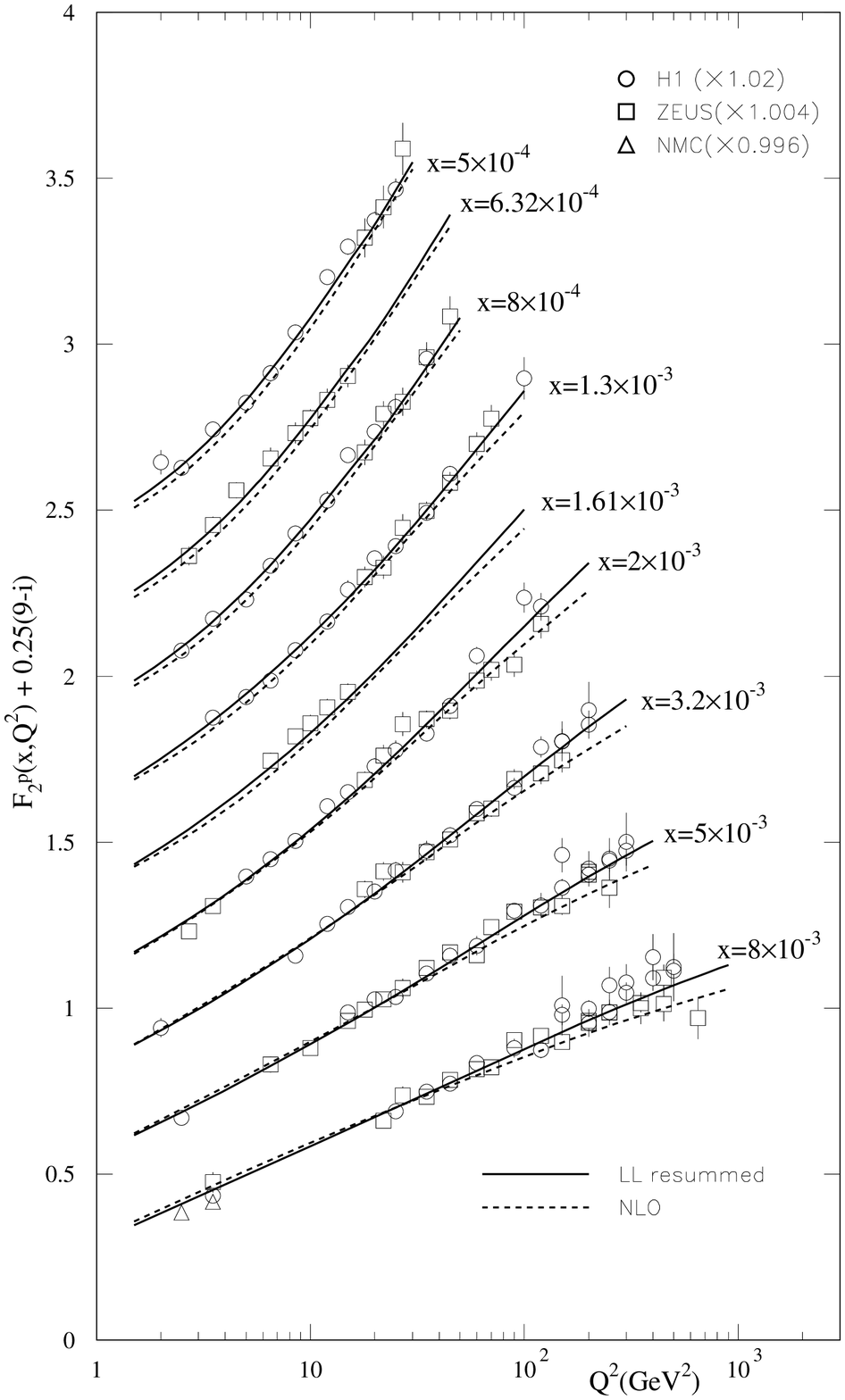}}
\caption{Theoretical predictions for the structure function $F_2$ alongside the data (normalisation dictated by the fit), for $5\times10^{-4}\leq x\leq8\times10^{-3}$.}
\label{data2}
\end{center}
\end{figure}

\begin{figure}
\begin{center}
\scalebox{0.7}{\includegraphics{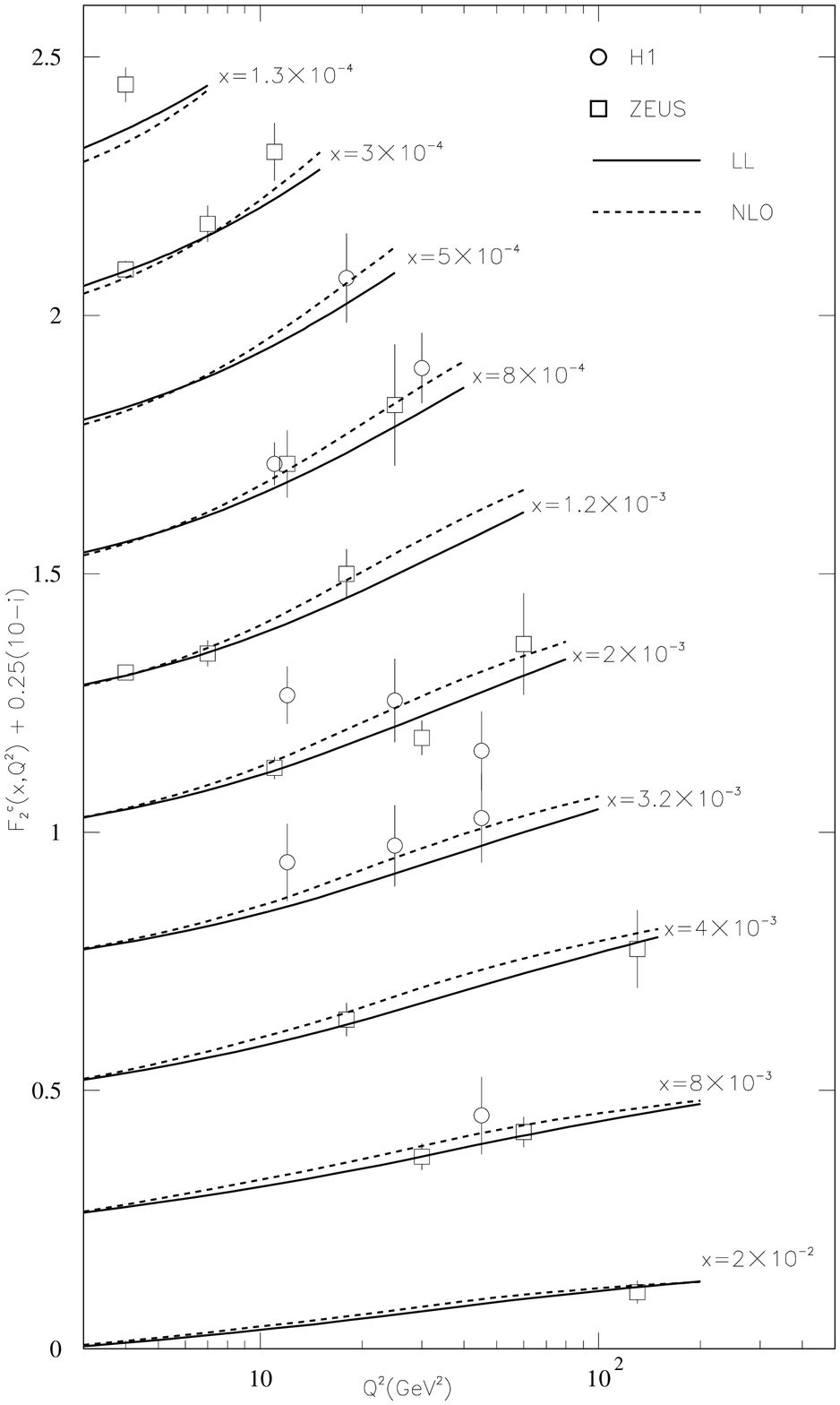}}
\caption{Resummed predictions for the charm structure function $F_2^c$ alongside HERA data, for $1.3\times10^{-4}\leq x\leq 2\times10^{-2}$.}
\label{charmdata}
\end{center}
\end{figure}

The description of the small $x$ DIS data is improved by resummation. In figures \ref{data} and \ref{data2} we show the resummed theoretical predictions alongside the data for a range of low $x$ values. The data are clearly fit very well, with no systematic tendency to undershoot. This is in contrast to results from a NLO fit, also shown in figures \ref{data} and \ref{data2}. The LL fit clearly has the increased slope needed to continue to fit the data well as $Q^2$ increases.
The charm structure function $F_2^c$ is also fit well, and in figure \ref{charmdata} we show the resummed theoretical prediction alongside the data and NLO fit results. Both the LL and NLO fits give a good overall fit, due to the somewhat large uncertainties associated with many of the data points. However, the resummed fit performs better at very small $x$. 
Even so, theory still underestimates $F_2^c$ at the lowest $x$ values ($\lesssim 3\times10^{-4}$). This may indicate the importance of higher orders in the resummed expansion. \\

Note that a poor fit is obtained for the DY data in the LL fit. This is not surprising, given the large perturbative corrections at NLO and the crude nature of the constant factor ${\cal K}^{DY}$. Consistent with this expectation, the NLO fit performs much better for this data set. The DY ratio data is better fit, which one expects given that some perturbative uncertainty cancels in taking the ratio of cross-sections. \\

Whilst the small $x$ fits to the HERA data improve, one can see from table \ref{LLfit2} that fits to some of the data sets actually worsen when resummations are implemented. One expects some tension between theory and data for non-electromagnetic DIS and other processes, due to the fact that resummed partons are being used without the appropriate resummed impact factors. However, these data sets do not contain points at very low values of $x$. Instead the problem lies at intermediate values of $x$ ($10^{-2}\lesssim x\lesssim 5\times10^{-1}$), where the theoretical prediction underestimates the deuterium data (proton structure function data, on the other hand, is fit rather well - in many cases the LL fit to $F_2^p$ outperforms the NLO fit). This indicates that the effect of the resummation in the partons is felt at higher values of $x$ than one na\"{i}vely expects. The higher evolution in the moderate $x$ region means that the sea quarks evolve much more quickly than the non-singlet quark combinations, leading to problems describing the relative shape of the sea and valence quarks. This accounts for the poorer fits to the deuterium and charged current data, and also the results for the Drell-Yan data. One sees from table \ref{LLfit2} that the E866 DY set is fit worse in the LL fit, even with the use of a variable K-factor. The exaggerated influence of small $x$ resummations is also evident in the value of $\alpha_S(M_Z^2)$ obtained in the LL fit of 0.1126, which is rather low compared to the world average. For comparison, the NLO fit gives $\alpha_S=0.120$ \footnote{One must bear in mind, however, that the definition of $\alpha_S$ is order dependent.}. \\

\begin{figure}
\begin{center}
\scalebox{0.9}{\includegraphics{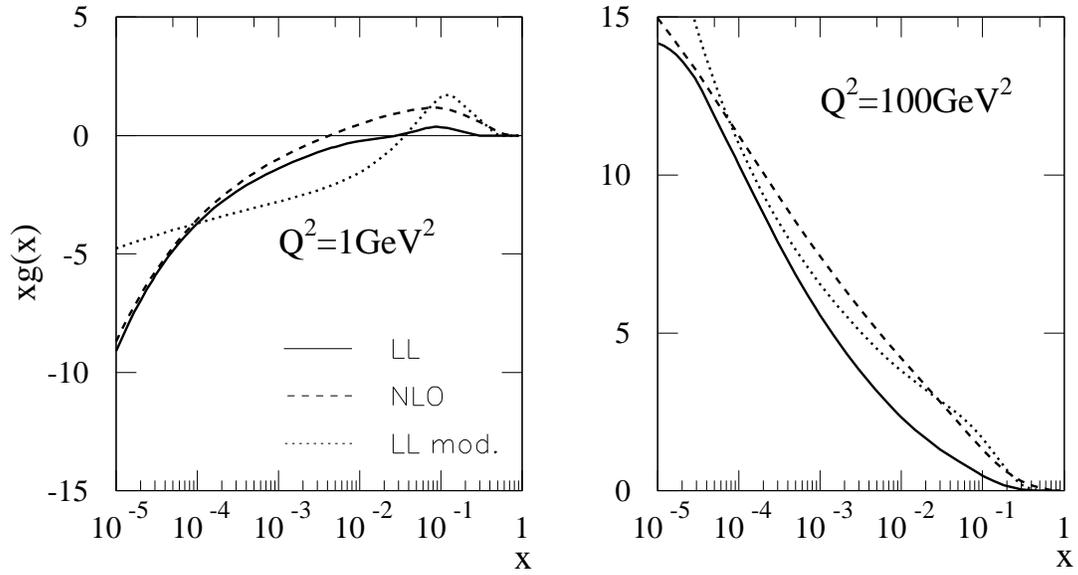}}
\caption{The gluon distribution at $Q_{init}^2=1\text{GeV}^2$ and $Q^2=100\text{GeV}^2$ obtained in the LL resummed fit and a previous NLO fit \cite{MRST2001}. Also shown is the result obtained from the modified LL fit described in the text.}
\label{gluons}
\end{center}
\end{figure}

We find that the gluon obtained from the LL fit is inconsistent with the gluon obtained from a previous NLO fit to the Tevatron jet data. The LL gluon is shown in figure \ref{gluons} together with a typical NLO gluon \cite{MRST2001} for two different scales.
At the starting scale in a LO fit, the gluon remains positive at small $x$, due to the fact that larger partons together with a raised coupling constant are needed to try and fit the HERA data. Once higher order corrections or resummations are implemented, one expects a lower gluon at small $x$, due to the faster evolution in this regime. In fact, figure \ref{gluons} shows that a negative gluon is needed at the starting scale in both the LL and NLO fits. However, the LL gluon turns negative at significantly higher $x$ (different by about an order of magnitude) than the NLO gluon. Furthermore, there is a large suppression of the LL result relative to the NLO gluon in the intermediate $x$ region. This is further evidence that the effects of the small $x$ resummations are felt too strongly at moderate $x$. The suppression of the gluon in the LL fit persists at higher $Q^2$, as can be seen from the right-hand panel of figure \ref{gluons}. We note that the left-hand plot shows the input scale gluon from the resummed fit agreeing extremely closely with the corresponding NLO gluon at small $x$. This is nothing more than a coincidence, as can be seen from the right-hand panel where the evolution drives the gluons to be quite different at higher scales.\\ 

The solution to the problem of resummations manifest at moderate $x$ is found by considering higher order terms in the fixed order expansion - NLO corrections help to suppress small $x$ resummation effects until lower values of $x$, due to the inclusion of the correct higher order moderate $x$ behaviour. Indeed, a NLO fit with no resummations shows significant improvement over a LO fit, and the subsequent improvement upon performing a NNLO analysis stems largely from the inclusion of the extra small and large $x$ terms (see \cite{MRSTNNLO}). The characteristic feature of the complete NLO $P_{qg}$ and $P_{qq}$ is a dip at moderate $x$, followed by a rapid increase at very high $x$ due to the presence of high $x$ divergences (absent in the $\msbar$ scheme). One can therefore approximate subleading behaviour at small $x$ by modifying the resummed splitting function, replacing the ${\cal O}(\alpha_S^2)$ term as follows:
\begin{equation}
\gamma_{qg}^{LL, mod.(2)}=34.67\left[\frac{1}{x}-\frac{A}{x^\alpha}\right],
\label{mod1}
\end{equation}
where one chooses the values of $A$ and $\alpha$ to closely model the dip at moderate $x$. We use $\alpha=0.47$ and $A=1.2$. Note that this does not have a serious impact on the splitting function at high $x$, given:
\begin{equation}
\lim_{x\rightarrow1}(x^{-1}-1.2x^{-\alpha})=-0.2,
\end{equation}
whereas the complete NLO DIS scheme splitting function is divergent as $x\rightarrow 1$. \\

We performed a global fit using this modified anomalous dimension (which also affects $P_{qq}$ by equation (\ref{colourcharge})). No modification to the resummed $P_{gg}$ (and, thus, $P_{gq}$ by the colour relation) was implemented. However, this has no term at ${\cal O}(\alpha_S^2)$ in the resummed expression. Furthermore, the NLO corrections in the fixed order quantity are small. We find a more favourable comparison to the gluon arising from the Tevatron jet data. The gluon which emerges is shown alongside those from the other fits in figure \ref{gluons} and one can see that although it is qualitatively the same as the LL result, the suppression at moderate $x$ is no longer as severe. Indeed, it is qualitatively more similar to the gluon obtained from a NLO or NNLO fit. This is therefore an indication that a next-to-leading order analysis (n.b. NLO and NLL) is what is really needed. It is still not possible even in the modified fit to obtain a very good fit to all the charged current data (or, for that matter, the gluon from the jet data), suggesting again the necessary inclusion of higher orders. 

\section{The Longitudinal Structure Function}
\begin{figure}
\begin{center}
\scalebox{0.8}{\includegraphics{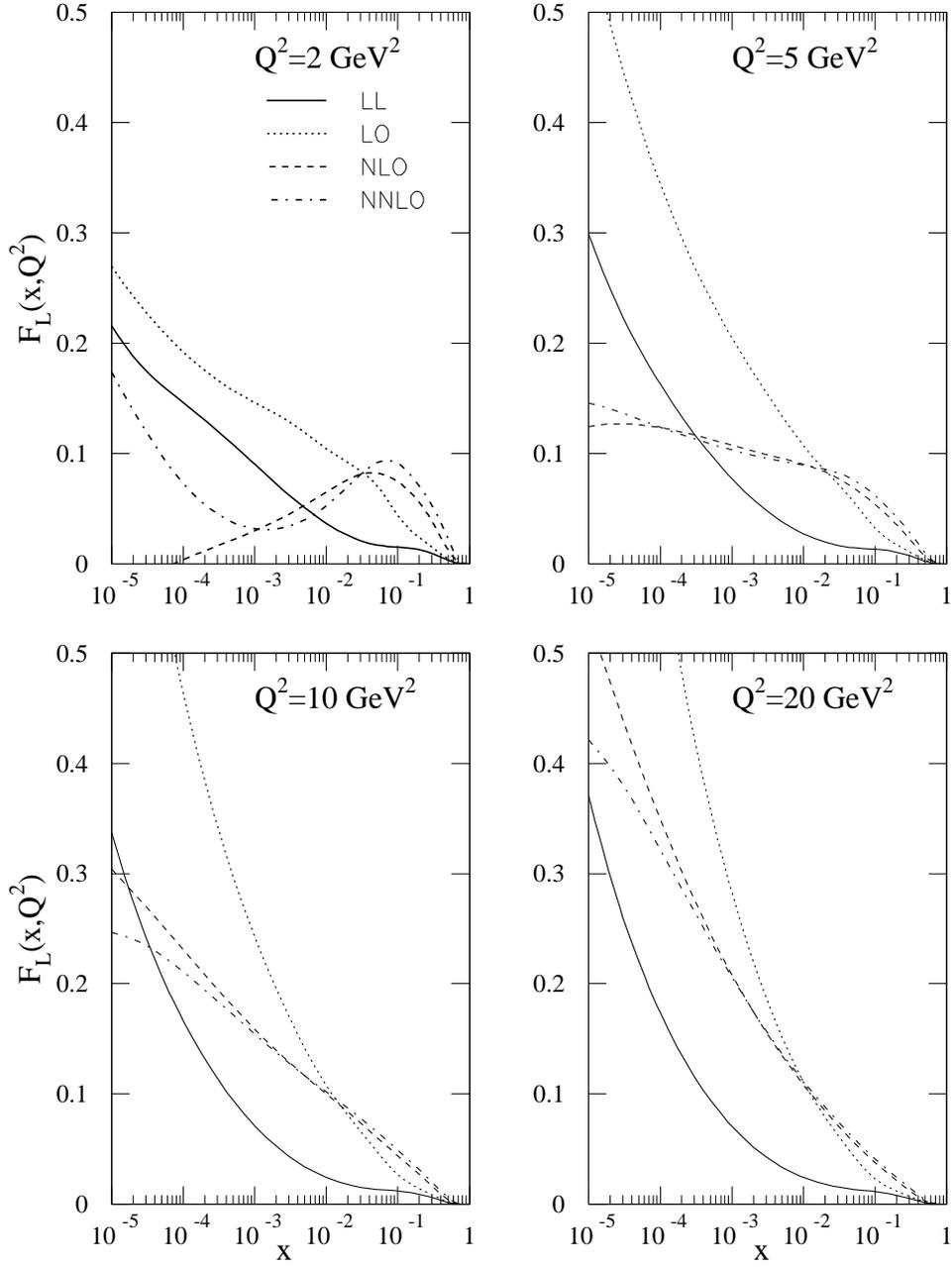}}
\caption{Resummed predictions for the longitudinal structure function contrasted with results from fixed order fits \cite{ThorneFL}.}
\label{flplot}
\end{center}
\end{figure}

In this section we discuss the resummed predictions for $F_L$. A plot of the value of the longitudinal structure function obtained from the LL resummed fit described in the previous section is shown for various values of $Q^2$ in figure \ref{flplot}. Also shown are the results obtained from the LO fit undertaken in this paper, together with similar results from previous NLO and NNLO fits \cite{MRSTfl}. One sees that the LO estimate is far higher at small $x$ when compared with higher orders in the fixed order perturbation theory, owing to the larger gluon at low $Q^2$ and the higher value of $\alpha_S$ needed to fit the $F_2$ data. Comparison with the NLO and NNLO curves reveals the apparent perturbative instability at small $x$. This is particularly noticeable in the $Q^2=2$GeV plot, corresponding to a scale at which the gluon is negative (at NLO and NNLO) at low $x$. One needs the divergent low $x$ terms at ${\cal O}(\alpha_S^3)$ in the longitudinal coefficient functions to counteract the negative small $x$ gluon leading to positivity of the structure function (see \cite{MRSTfl} for a discussion). The poor convergence of the fixed order calculations at low $x$ (of greater significance at low $Q^2$) indicates that high energy resummations are probably necessary in this regime. \\

One sees that at small $x$ and $Q^2$, the resummed result lies above the NNLO fit, but evolves more slowly than the fixed order fits as $Q^2$ increases. The important point to note is that it is much more stable than the fixed order results at small $x$, and lies somewhere between the LO and NLO results. A less desirable feature of the resummed fit, however, is the behaviour at moderate and high $x$. In these regimes the resummed result is severely below the fixed order results, even as $Q^2$ reaches values at which the NLO and NNLO results are closer together (n.b. denoting less sensitivity to the small $x$ instability). The reason for this, as noted in the previous section, is that the gluon from the resummed fit is much smaller at higher $x$ than should really be the case. Also, the lower value of $\Lambda$ generated by the resummed fit decreases the contributions from the quarks at higher $x$. \\

\begin{figure}
\begin{center}
\scalebox{0.8}{\includegraphics{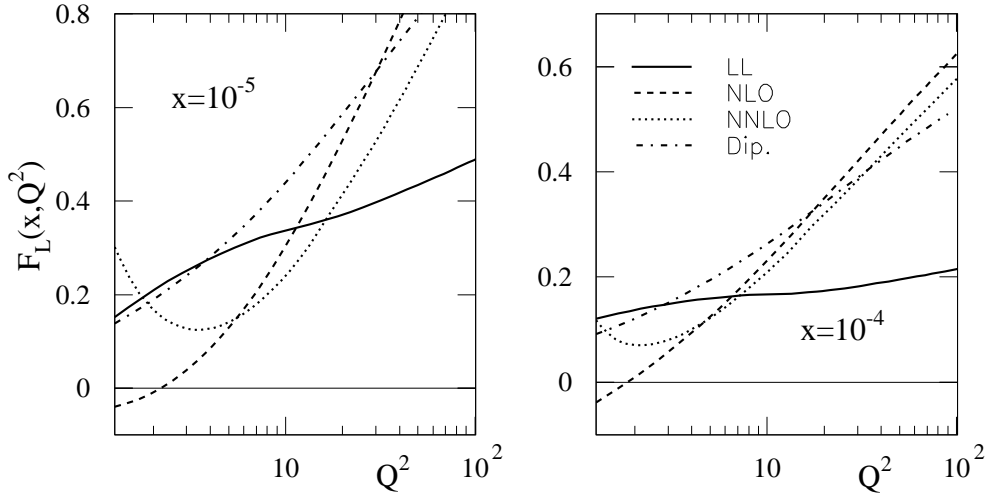}}
\caption{Resummed prediction for the longitudinal structure function as $Q^2$ varies, for $x=10^{-5}$ and $x=10^{-4}$. Also shown are fixed order results, and the result from the dipole fit of \cite{Thornedipole}.}
\label{flx}
\end{center}
\end{figure}

To emphasise the improvement at low $x$ values in the resummed fit, we show in figure \ref{flx} the evolution of $F_L$ for $x=10^{-5}$ and x=$10^{-4}$.
The instability in the fixed order results is clearly visible at lower $Q^2$, where the resummed fit is much more sensible. The resummed result is much flatter in $Q^2$ than the fixed order results.\\

\begin{figure}
\begin{center}
\scalebox{0.8}{\includegraphics{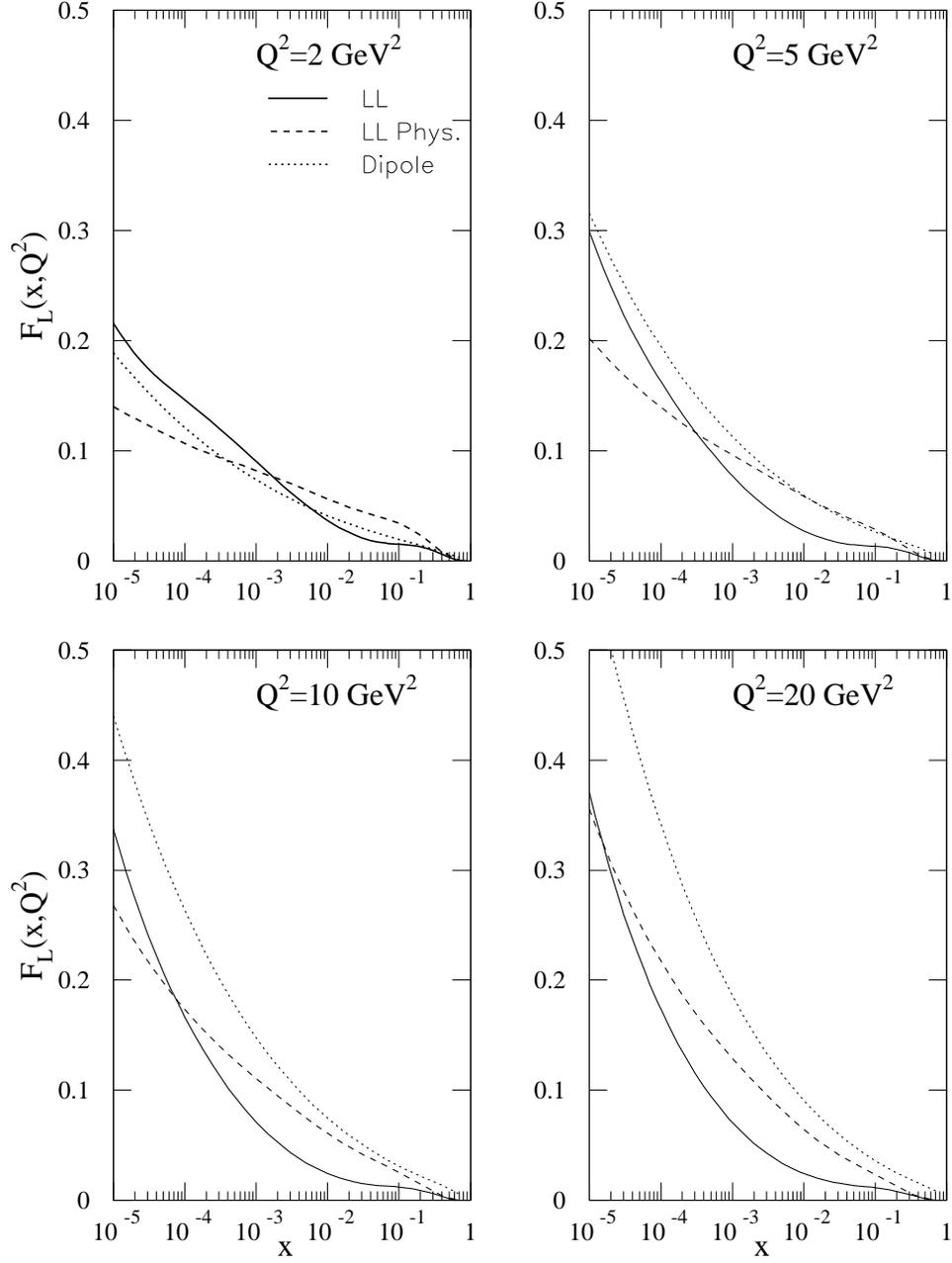}}
\caption{Resummed prediction for $F_L$ alongside the result obtained from a previous resummed fit using physical anomalous dimensions. Also shown is the structure function obtained from the dipole model fit of \cite{Thornedipole}.}
\label{fl2plot}
\end{center}
\end{figure}

It is interesting to compare the resummed fit performed here where the evolution is set up in terms of parton densities, to a previous resummed fit which evolves the structure functions directly \cite{ThorneFL} using physical anomalous dimensions \cite{Catani97} and which had a significantly less rigorous treatment of heavy flavours. From figure \ref{fl2plot} one sees that the results are qualitatively consistent between the two fits. They converge at very high $x$ to a value lower than the fixed order results, given that they both suffer from a gluon which is too small in this region. Also, NLO and NNLO corrections to the longitudinal coefficient functions are important at high $x$, and these are missing from the resummed fit. However, the evolution is moderated a little at intermediate $x$ in the physical anomalous dimension fit, due to an additional convolution in evaluating the derivative of the structure functions. For example, one has for the singlet structure function:
\begin{align}
\frac{\partial F_2(x,Q^2)}{\partial\log{Q^2}}&=\Gamma_{2L}(x)\otimes F_L(x,Q^2)+\Gamma_{22}(x)\otimes F_2(x,Q^2)\\
&=\Gamma_{2L}(x)\otimes \left[C_{Lg}\otimes g(x,Q^2)+C_{Lq}\otimes\Sigma(x,Q^2)\right]+\ldots
\label{flphys}
\end{align}
where the ellipsis denotes the term in $F_2$. The term in the gluon distribution thus involves an intermediate convolution with the coefficient:
\begin{equation}
C_{Lg}(x)=\frac{\alpha_S}{4\pi}4x(1-x) +{\cal O}(\alpha_S^2).
\label{clgm0}
\end{equation}
This tends to zero as $x\rightarrow 1$, and the effect of this additional convolution is to suppress the phenomenological effect of the small $x$ resummations in the physical anomalous dimension $\Gamma_{2L}$ to lower values of $x$. The result is that the effective gluon and $F_L$ in such a fit are increased in the moderate $x$ region of $x\sim10^{-1}-10^{-2}$. Both the resummed fits are lower than the fixed order results at moderate $x$, as can be expected from the suppressed gluon. \\

In figure \ref{fl2plot}, we also show the value of $F_L$ obtained from the dipole model fit presented in \cite{Thornedipole}, itself similar in approach to previous dipole fits \cite{Golec-Biernat,Forshaw,Iancu,Machado}. There, the structure functions are written as the convolution of a dipole cross-section (dependent on the unintegrated gluon distribution) and a dipole wavefunction representing the probability of a photon fluctuating into a quark-antiquark pair. This is essentially a restatement of the LL $k_T$ factorisation theorem of equation (\ref{highfactor}), although the gluon distribution is modelled differently using a parameterisation based on evolution according to the approximate anomalous dimension:
\begin{equation}
\gamma_{gg}^{dip.}=\bar{\alpha}_S(Q^2)\left(\frac{1}{N}-1\right).
\label{dipgg}
\end{equation}
One sees from the figure that at low $Q^2$ the dipole fit is very close to the LL resummed fit performed here. They both share the feature of a gluon which is much too small, compared to a DGLAP gluon, at moderate and high $x$ - where the DGLAP results should be reliable. As $Q^2$ increases, however, the dipole result overshoots the resummed fit, owing to its much higher evolution. This can also be seen in figure \ref{flx}, where the behaviour of the dipole result as $Q^2$ increases is shown. It must be stressed that both the dipole and LL resummed fits give very good descriptions of the low $x$ data. The shortcoming of both, however, is that they are not quantitatively correct over the whole $x$ range, owing to the need to suppress the small $x$ resummation effects at higher $x$. Note that the $k_T$ factorisation approach is easier to extend to higher orders. At LL order it is formally equivalent to the dipole approach. Beyond LL order, the dipole approach suffers conceptual problems. For example, including the exact gluon kinematics in the photon-gluon impact factor (a significant effect at NLL and higher orders) leads to the non-conservation of impact parameters describing transverse dipole size throughout the interaction \cite{Peschanski}, which is a necessary property for the applicability of dipole arguments. Although it is possible this property may be recovered by inclusion of full NLL effects in the dipole wavefunction, there are no such conceptual barriers to extending the $k_T$ factorisation approach. Furthermore, there are no difficulties in matching to the fixed order expansion in the factorisation framework. It is also more suited to the consideration of proton-proton collisions, which do not have a direct interpretation in terms of dipoles. 

\section{Conclusions}
In this paper we have proposed a new variable flavour number scheme, the $\DIS(\chi)$ scheme, for use in deep inelastic scattering. The motivation for this scheme is to provide a way of carrying out heavy flavour calculations at small $x$ in a form that can be consistently implemented alongside the fixed order QCD expansion. We have shown how to obtain the resummed coefficients and heavy matrix elements at small $x$ in double Mellin space from existing expressions for the photon impact factors at LL order. One then obtains results in $x$ and $Q^2$ space after solving the BFKL equation. Here we have adopted a particular solution method which resums running coupling corrections, but the formulation of the $\DIS(\chi)$ scheme in $(N,\gamma)$ space is more general than this and can be used with other BFKL approaches at small $x$. Alternatively, it can be used at fixed order even if one is not considering high energy resummation.\\

The advantage of the method adopted here is that it produces analytic results which can be directly used in parton fitting routines. Furthermore, $\bb$-dependent corrections are effectively resummed to all orders (at least, to any desired phenomenological precision). It offers a more physically motivated definition of both the massless partons (due to the use of the DIS scheme), and the heavy partons (due to the choice of heavy quark coefficients). We used our scheme in a LO global fit to scattering data, including LL resummations with running coupling corrections. The running coupling is seen to significantly moderate the strong high energy divergence observed in physical quantities in a purely LL fit. A good overall fit is obtained, and the fit to HERA data at small $x$ is improved relative to the LO fit with no resummations. The description of the HERA charm data for $F_{2}^c$ is also improved. Compared with a NLO fit, the LL results give a significantly better description of the low $x$ data and describe most of the proton data well. However, the fit worsens at moderate and high $x$ for the non-proton data, due to the fact that the small $x$ resummations lead to a decrease of the coupling constant, and an incorrect relative shape between the sea and valence quarks due to the increased evolution of the former. There is also a very small gluon distribution in the moderate $x$ region, which is inconsistent with previous gluons obtained from Tevatron jet data. This suggests that higher order corrections in both the fixed order and resummed expansions are what is really needed to understand the data. The effect of higher order corrections is to suppress the influence of small $x$ resummations to lower values of $x$. We demonstrated this here by making a small modification to the LL fit, which tempers the increased evolution of the quarks at moderate $x$. This resulted in a gluon distribution more in line with that obtained from the Tevatron jet data. The gluon from the modified fit shows a significant increase in the moderate $x$ region, as expected, which allows for a better description of the high $x$ data.\\

One does of course expect some degree of tension in the resummed fit between the theoretical predictions and the data for non-DIS sets. This is due to the use of partons obtained using small $x$ resummed evolution kernels, but without including resummed coefficients for e.g. charged current scattering. This should not be too significant in practice, however, given that the non-DIS data sets do not include data points at very low $x$. The fact that the non-DIS data are not well fit is itself an indication that NLO corrections are needed.\\

Our resummed results for the longitudinal structure function were presented in section 5. Importantly, the resummed fit produces much more stable results at low $x$ than the fixed order results, particularly at low $Q^2$ where resummations are expected to be more important due to the higher coupling constant. However, at moderate and high $x$ the resummed results undershoot the fixed order results. This feature is shared at lower $Q^2$ by the dipole model fit of \cite{Thornedipole}, which together with the LL fit leads to a gluon which is much too small at moderate $x$. A previous resummed fit using physical anomalous dimensions gives similar qualitative results. One does not expect to match the NLO and NNLO results at high $x$ in the LL fit, due to missing high $x$ terms in the longitudinal coefficients. The results at low $x$ are very encouraging, however, and give a strong indication that high energy resummations are a necessary prerequisite for a sensible prediction of $F_L$.\\

Our final conclusion is therefore that, although we have demonstrated the need for high energy resummations in order to correctly describe structure function data at small $x$, higher order corrections in the fixed order expansion are needed to achieve a truly quantitative agreement over the full $x$ range. For consistency one should build NLL resummations over a NLO fixed order expansion. Our investigation into implementing such a scheme in an approximate framework is ongoing \cite{White}.\\
\section{Acknowledgements}
CDW wishes to thank Jeppe Andersen for many helpful discussions, and is grateful to PPARC for a research studentship. RST thanks the Royal Society for the award of a University Research Fellowship. \\

\bibliography{refs}
\end{document}